\pdfoutput=1
\documentclass[twocolumn]{aastex701}
\usepackage{bm}           
\usepackage{amsmath}
\usepackage{braket}

\newcommand{\divr}{\nabla\cdot}

\begin{document}

\title{The Curious Case of V CVn}
\shorttitle{The Curious Case of V CVn}
\shortauthors{Power et al.}

\author[0009-0005-1455-8093]{Michael T. Power}
\affiliation{David A. Dunlap Department of Astronomy \& Astrophysics, University of Toronto, 50 St. George Street, Toronto, ON M5S 3H4, Canada}
\affiliation{Department of Physics and Physical Oceanography, Memorial University of Newfoundland, 283 Prince Philip Drive, St. John's, NL A1B 3X7, Canada}
\email[show]{michael.power@astro.utoronto.ca}  

\author[0000-0002-7322-7236]{Hilding R. Neilson} 
\affiliation{Department of Physics and Physical Oceanography, Memorial University of Newfoundland, 283 Prince Philip Drive, St. John's, NL A1B 3X7, Canada}
\email{hneilson@mun.ca}

\author[]{Galina Sherren}
\affiliation{Department of Physics and Physical Oceanography, Memorial University of Newfoundland, 283 Prince Philip Drive, St. John's, NL A1B 3X7, Canada}
\email{gsherren@mun.ca}

\begin{abstract}
Decades of observations on the star V Canum Venaticorum (V CVn) have revealed an unusual inverse relationship between its linear polarization and light curves (sometimes with a lead/lag time between them) and an almost constant polarization position angle. One theory proposed to explain this behaviour is the existence of a bow shock driven by a spherically symmetric time-varying dusty wind from the star, which is assumed to vary due to radial pulsations. To test this hypothesis, this study uses a new framework developed in \textsc{ZEUS3D}, a multiphysics magnetohydrodynamics code. The results of this work show that when a time-varying stellar wind is at its maximum brightness, the polarization signal is at a minimum due to the wind structure and a dense, symmetric shell that forms around the star. Conversely, when the brightness is at a minimum, the symmetric shell around the star is much less dense, and the polarization is instead dominated by the asymmetric bow shock structure, causing the polarization signal to attain a maximum value. Numerically reproducing the observed inverse relationship between the polarization and light curve provides a strong theoretical argument that a variable stellar wind bow shock is the solution to the curious case of V CVn.
\end{abstract}
\section{Introduction}
V Canum Venaticorum (V CVn) is a semi-regular variable star in the constellation Canes Venatici, located at approximately $501\,\mathrm{pc}$ from Earth (parallax $1.9945 \pm 0.1175\,\mathrm{mas}$ (see, e.g.\@\citet{2020yCat.1350....0G}). Its Galactic coordinates are roughly $l \simeq 107.89^\circ$ and $b \simeq 70.77^\circ$. Classified as an M4--M6e star \citep{1937ApJ....85....9J}, V CVn exhibits significant variability and strong circumstellar dust scattering, resulting in an observed maximum polarization of approximately 6\% (see, e.g. \citet{Neilson2014}). Its proper motion is roughly $-38.13\,\text{mas}\,\text{yr}^{-1}$ in right ascension and $-12.11\,\text{mas}\,\text{yr}^{-1}$ in declination (see, e.g.\@\citet{2020yCat.1350....0G}); when combined with its distance, these values imply a tangential velocity on the order of $100\,\text{km}\,\text{s}^{-1}$. In contrast, its radial velocity is only around $-4.70\,\text{km}\,\text{s}^{-1}$ \citep{2009A&A...498..627F}, making the high tangential speed the dominant component of its space motion. This unusually high velocity indicates that V CVn is a runaway star \citep{Wenger_2000,2016A&A...595A...1G,2023A&A...674A..32B}.

The curious behaviour of V CVn is detailed in \cite{Neilson2014} and \cite{Neilson2023}. They take multiple decades of observations \citep{Wolff1996, Magalhaes1986, Poliakova1981} for V CVn and show that the brightness and polarization have a roughly inverse relationship, with a slight lead or lag time. When V CVn reaches its maximum brightness, the polarization nears its minimum, and when the brightness reaches a minimum, the polarization nears its maximum value. \cite{Neilson2014} and \cite{Neilson2023} noticed that the polarization position angle is roughly constant, which causes them to claim that this must imply the existence of an asymmetric structure which is mostly stable in time. They argue that since V CVn is a runaway star (see, e.g.\@\citet{2016A&A...595A...1G, 2023A&A...674A...1G, 2023A&A...674A..32B}) the presence of a stellar wind bow shock is likely, which is an asymmetric, mostly stable structure that can cause a large polarization signal to manifest \citep{Shrestha2018, Shrestha2021}. 

Some other proposals attempt to explain the curious behaviour of V CVn, such as \cite{Safonov2019}, which use speckle interferometry to claim that multiple `blobs' or arcs around the star are lit sequentially by non-radial pulsations from V CVn. This would mean that when the star is at maximum brightness on the far side, the Earth observes minimum brightness, but the blob behind the star, which will see maximum brightness, will scatter a maximal amount of polarized light toward the Earth. This would imply maximum polarization at minimum brightness. The reverse situation would occur when the star's far side is at minimum brightness, but the close side is at maximum brightness. The Earth would observe maximum brightness, but a minimum amount of polarized light would be scattered. 

Although the blob theory is appealing, the speckle interferometry may not be sensitive to a large-scale bow shock structure, which \cite{Neilson2023} argue likely exists due to V CVn's runaway status, similar to many other runaway stars. There are a handful of other theories involving magnetic fields, rapid rotation, etc., but \cite{Neilson2023} argue that these scenarios are also unlikely due to the nature of V CVn. Thus, this work aims to numerically test the hypothesis that a spherically symmetric wind, which varies with time, is the cause of the inverse relationship between the polarization and brightness of V CVn. 

This work aims to address two problems. The first is the computation of the full range of hydrodynamics in an interaction between the stellar wind from V CVn and the ISM it is moving through. The second is the calculation of the polarization signal directed toward Earth.

The paper is organized as follows: Section \ref{sec:Simulations_and_Models} describes the numerical methods, initialization of the problem and boundary conditions, and our models and corresponding parameters. Section \ref{sec:Results_and_Discussion} details the results of the numerical simulations and discusses the hydrodynamics and the corresponding polarization signal and correlations. Finally, Section \ref{sec:Conclusions} summarizes our interpretation of the results of this work and lists possible avenues for exploration in the future. 

\section{Code Description and Simulation Models}\label{sec:Simulations_and_Models}
\textsc{ZEUS3D} is a fully conservative computational magnetohydrodynamics grid code that uses a staggered mesh and an operator splitting scheme for numerical stability \citep{Clarke1996ApJ...457..291C} and accounts for many physical phenomena, including viscosity, gravity, molecular cooling, and ambipolar diffusion; these are just its single fluid capabilities \citep{clarke2016zeus3d}. The equations which \textsc{ZEUS3D} solves for our problem are as follows,
conservation of mass:
\begin{equation}
    \label{Continuity}
    \frac{\partial\rho}{\partial t} + \nabla \cdot (\rho\bm{v}) = 0;
\end{equation}
conservation of momentum:
\begin{equation}
    \frac{\partial \bm{s}}{\partial t} + \nabla\cdot \biggl[\bm{s}\bm{v} + p\bm{1} - \mu \bm{S}\biggr] = \bm{0};
\end{equation}
and the choice of either conservation of internal energy, which keeps pressures numerically positive-definite, but total energy may not be conserved to machine round off error \citep{Clarke_2010}:
\begin{equation}
    \frac{\partial e}{\partial t} + \nabla \cdot (e\bm{v})=-p\divr\bm{v}+\mu\bm{S}\mathbin{:}\nabla\bm{v};
\end{equation}
or the conservation of total energy, which conserves total energy to machine round off error but may cause negative pressures to manifest \citep{Clarke_2010}:
\begin{equation}
\frac{ \partial e_{\text{T}} }{ \partial t } + \divr\biggl[ (e_{\text{T}}+p)\bm{v} -\mu\bm{S}\cdot\bm{v} \biggr]=0.
\end{equation}

Here, $\rho$ represents the mass density of the fluid, $t$ is the time, $\bm{v}$ is the velocity field, $\bm{s}=\rho\bm{v}$ is the momentum field, $p$ is the thermal pressure, $\bm{1}$ is the identity tensor, $\mu$ is the viscosity \citep{VonNeumann1950}, $\bm{S}=\partial_jv_i+\partial_iv_j-(2/3)\delta_{ij}\divr\bm{v}$ is the shear tensor \citep{Stone1992ApJS...80..753S}, $e$ is the internal energy density, and $e_{\text{T}}=e+\rho v^2/2$ is the total energy density.  

The system of partial differential equations must then be closed with an equation of state to relate the thermodynamic variables. For many applications in astrophysics, the ideal gas law is an excellent approximation of reality. Therefore, \textsc{ZEUS3D} uses it to close the system (although other equations of state may be employed if desired). Thus, the ideal gas law is
\begin{equation}\label{IdealGasLaw}
    p=(\gamma-1)e,
\end{equation}
where $\gamma = 5/3$ is the ratio of specific heats. 

Within \textsc{ZEUS3D}, we have implemented routines to calculate the observed polarization signal, assuming the case of optically thin Thomson scattering. The implementation of the polarization calculations in \textsc{ZEUS3D} is quite straightforward. If one works in two dimensions assuming azimuthal symmetry, \cite{Brown1977} gives the polarization signal, which we have cast into a more useful form for \textsc{ZEUS3D} as 
\begin{equation}\label{Gammapolarization}
    \gamma_{\text{p}} = \frac{\int_0^\infty dr\,\int_0^\pi d\theta\, \rho(r,\theta)\cos^2(\theta)\sin(\theta)}{\int_0^\infty dr\,\int_0^\pi d\theta\, \rho(r,\theta)\sin(\theta)},
\end{equation}
\begin{equation}\label{OpticalDepth}
    \bar{\tau} = \frac{\pi\sigma_0}{\bar{m}}\int_0^\infty dr\,\int_0^\pi d\theta\,\rho(r,\theta)\sin(\theta),
\end{equation}
\begin{equation}\label{Residualpolarization}
    P_{\text{R}}\simeq \bar{\tau}(1-3\gamma_{\text{p}})\sin^2(i).
\end{equation}
Here, $\gamma_{\text{p}}$ is a geometric factor which governs the polarization due to scattering in a density field with azimuthal symmetry, $\rho(r,\theta)$, $\bar{m}$ is the average particle mass, $\sigma_0$ is related to the wavelength independent Thomson scattering cross section, $\sigma_{\text{T}}=6.66\times 10^{-25}\,\text{cm}^2$, by $\sigma_0=3\sigma_{\text{T}}/(16\pi)$, $\bar{\tau}$ is effectively the optical depth of the envelope averaged over the angular coordinate, and $P_{\text{R}}$ is the residual polarization at an angle of inclination, $i$ defined by \cite{Brown1977}. Thus, in two dimensions, \textsc{ZEUS3D} will write the values of the geometric factor, $\gamma_{\text{p}}$, to a file at each time step after the hydrodynamic cycle completes. If one works in three dimensions, \textsc{ZEUS3D} will write the values of the normalized Stokes parameters $Q$ and $U$ defined by \cite{Brown1978} to a file at each time step after the hydrodynamic cycle completes. For the numerical integration of the density field within \textsc{ZEUS3D} we have implemented is accomplished by choice of either Romberg integration \citep{press2003numerical} including a Richardson Extrapolation step \citep{richardson1911approximate, richardson1927deferred}, Monte Carlo integration \citep{press2003numerical}, or a simple Riemann sum.

Since V CVn is a semi-regular variable star, we model the mass-loss rate -- taken as a proxy for the brightness -- as a function which varies with a period equivalent to the observational data of the star. Since V CVn is an AGB star, one may use the relation from \cite{DeBeck} for an average mass-loss rate. The relationship defines the average mass loss of an AGB star, $\dot{\bar{M}}$, (in solar masses) as a function of its period, $\tau$, (in days) as
\begin{equation}
    \log(\dot{\bar{M}})= \begin{cases}
        -7.37 + 3.42\times10^{-3}\tau, & \tau \lesssim 850, \\
        -4.46, & \tau \gtrsim 850.
    \end{cases}
\end{equation}
%\subsection{Initial and Boundary Conditions}
To model V CVn in \textsc{ZEUS3D}, we follow the work of \cite{MacKey2021}, where a `wind bubble' resolves the wind coming from the star. Essentially, for the wind bubble approximation, one calculates the average standoff distance to the bow shock \citep{baranov1971, Wilkin1996}, 
\begin{equation}
    \bar{R}_{\text{SO}} = \sqrt{\frac{\dot{\bar{M}}\bar{v}_{\text{w}}}{4\pi\rho_0v_*^2}}.
\end{equation}
Here, $\bar{v}_{\text{w}}$ is the average speed of the stellar wind, $\rho_0$ is the density of the ISM, and $v_*$ is the star's speed; equivalently, this is the speed of the ISM in the rest frame of the star. This sets the overall scale of the problem, and one then determines an appropriate minimum scale for the problem, which is some factor, $\zeta$, less than the standoff distance to the bow shock. One chooses this factor, $\zeta$, large enough to have sufficient computational zones between the wind bubble -- which now acts as a boundary condition internal to the grid -- and the bow shock to resolve the necessary physics. 

Coupled with the wind bubble, we introduce a static mesh refinement scheme to save computational resources while ensuring an adequate level of resolution to capture all physics sufficiently. We work in two-dimensional spherical-polar coordinates under the assumption of azimuthal symmetry and take the z-axis (pointing from left to right in our simulations) as the star's direction of motion. We perform the numerical simulations within the star's rest frame, use pure ideal hydrodynamics, and resolve the simulation on a $570\times720$ grid with $20$ zones of static mesh refinement in the radial direction to adequately capture the wind bubble. The initial setup of the ISM is uniform, and we treat the wind bubble as a boundary condition where all values within are held constant or vary with time in a predetermined way.

\subsection{Static Wind Velocity Model}
\label{sec:static_model}
Since we do not model the stellar wind directly by determining a model and then solving the governing equations for its time evolution (e.g. see \cite{henny_stellar_winds}), a periodic prescription is assumed. For this work and the case of a static wind velocity, the assumed form of the time-varying mass-loss rate of the wind is 
\begin{equation}\label{GeneralMAssLossSW}
    \dot{M}(t)= \alpha + \frac{\beta}{2} [1-\cos(\omega t)].
\end{equation}
Here, we define the oscillation frequency in the usual way, $\omega=2\pi/\tau$; both $\alpha$ and $\beta$ represent constants. Using $1-\cos(\omega t)$ allows the mass-loss rate to vary to any degree one desires while ensuring it is positive-definite. In a case where $\sin(\omega t)$ alone governs the variation in the mass-loss rate, it has a limit of twice its average, and this limit would set the minimum mass-loss rate to zero. Two more parameters are required to determine the primitive variables uniquely. For this case, the fixed speed of the stellar wind $\bar{v}_\text{w}$ will act as the first, and the second will be the level to which the mass loss varies, defined by the ratio
\begin{equation}\label{DefinitionOfEta}
    \eta = \frac{\dot{M}_\text{Max}}{\dot{M}_\text{Min}}.
\end{equation}
Using these conditions, one may determine a unique solution for the parameters $\alpha$ and $\beta$, as well as the remaining primitive variables for the initialization of the simulation. After some algebra, one may find that
\begin{equation}\label{MassLossRateStaticSolved}
    \dot{M}(t)=2\cdot\frac{1+(\eta-1)\sin^2(\tfrac{\omega}{2} t)}{\eta + 1}\dot{\bar{M}},
\end{equation}
\begin{equation}\label{StaticWindDensitySolved}
    \rho_\text{w}(t)=2\cdot\frac{1+(\eta-1)\sin^2(\frac{\omega}{2} t)}{\eta + 1}\cdot\frac{\dot{\bar{M}}}{4\pi R_\text{w}^2\bar{v}_\text{w}},
\end{equation}
and
\begin{equation}\label{WindInternalEnergySolved}
    e_\text{w}(t)=\frac{2k_\text{B}T_\text{w}[1+(\eta-1)\sin^2(\tfrac{\omega}{2} t)]}{(\eta+1)(\gamma-1)\bar{m}}\cdot\frac{\dot{\bar{M}}}{4\pi R_\text{w}^2\bar{v}_\text{w}}.
\end{equation}
Table \ref{tab:Variable_Params} lists this model's assumed and calculated variable values.
\begin{table}
    \centering
    \caption{Simulation Parameters}
    \label{tab:Variable_Params}
    \begin{tabular}{lll}
%        \toprule
        \multicolumn{3}{c}{\textbf{Stellar Wind Parameters}} \\
%        \midrule
        \textbf{Variable} & \textbf{Value (CGS)} & \textbf{Meaning} \\
%        \midrule
        $\dot{\bar{M}}$ & $1.26\times 10^{19}$ & Average mass-loss rate \\
        $\bar{v}_\text{w}$ & $2.50\times 10^{6}$ & Average Wind Speed \\
        $\tau$ & $1.68\times 10^{7}$ & Period \\
        $\omega$ & $3.73\times 10^{-7}$ & Frequency \\
        $T_\text{w}$ & $3.20\times 10^{3}$ & Wind Temperature \\
        $R_\text{w}$ & $2.00\times 10^{15}$ & Wind Bubble Radius \\
        $\eta$ & $10$ & Mass-loss rate Ratio \\
        $\lambda$ & $3$ & Wind Speed Ratio \\
        $\zeta$ & $12.9$ & Wind Bubble Ratio \\
%        \midrule
        \multicolumn{3}{c}{\textbf{ISM Parameters}} \\
%        \midrule
        \textbf{Variable} & \textbf{Value (CGS)} & \textbf{Meaning} \\
%        \midrule
        $v_*$ & $1.50\times 10^{7}$ & ISM Speed \\
        $n_0$ & $1.00\times 10^{1}$ & ISM Number Density \\
        $\rho_0$ & $1.67\times 10^{-23}$ & ISM Density \\
        $T_0$ & $2.50\times 10^{1}$ & ISM Temperature \\
        $e_0$ & $5.18\times 10^{-14}$ & ISM Internal Energy \\
%        \midrule
        \multicolumn{3}{c}{\textbf{Other Parameters}} \\
%        \midrule
        \textbf{Variable} & \textbf{Value (CGS)} & \textbf{Meaning} \\
%        \midrule
        $\bar{m}$ & $1.67\times 10^{-24}$ & Mean Particle Mass \\
        $\bar{R}_\text{SO}$ & $2.58\times 10^{16}$ & Mean Standoff Distance \\
        $R_\text{Sim}$ & $5.00\times 10^{17}$ & Radius of Simulation \\
        $t_\text{Sim}$ & $7.00\times 10^{11}$ & Total Simulation Time \\
%        \bottomrule
    \end{tabular}
    \begin{tabular}{ll}
%        \toprule
        \multicolumn{2}{c}{\textbf{Time-Varying Parameters}} \\
%        \midrule
\textbf{Variable} & \textbf{Functional Form (CGS)}\\
%        \midrule
        $\dot{M}(t)$ & $(2.29\times 10^{18})+(2.06\times 10^{19})\cdot\sin^2(\omega t)$ \\
        $\rho_\text{w}(t)$ & $(1.82\times 10^{-20})+(1.64\times 10^{-19})\cdot\sin^2(\omega t)$ \\
        $e_\text{w}(t)$ & $(7.22\times 10^{-9})+(6.50\times 10^{-8})\cdot\sin^2(\omega t)$ \\
%        \bottomrule
    \end{tabular}
\end{table}

\subsection{An Effective Theory of Mie Scattering}
While Thomson scattering is a decent first approximation and provides meaningful insights into how the roughly inverse relationship between brightness and polarization emerges, the signal is very weak and does not rise to the observed strength. Realistically, the environment around V CVn will be dusty, meaning that the scattered light we measure on Earth will likely be due to dust grains. To account for this, we develop an averaged theory of Mie scattering.

We begin by defining some differential contributions in the form of Stokes parameters from each cell in the computational domain. We take the differential contribution for the Stokes parameter, $I$, to be the product of the stellar intensity incident on the particular volume element, the average scattering cross-section per volume to account for how much of the light will scatter, the phase function to determine what fraction of light scatters into each angle, and the differential volume of the element itself. Mathematically, we can write this as
\begin{equation}
    dI_\text{scat} = [I_*]\times[n_\text{dust}\braket{\sigma_\text{scat}}]\times[\Phi]\times[dV].
\end{equation}
Here, 
\begin{equation}
    I_* = \frac{L_*}{4\pi r^2},
\end{equation}
is the intensity of the starlight incident upon a cell in the computational domain at radius $r$. Note that the luminosity of the star, $L_*$, may need to be taken at the retarded time and, therefore, may have an underlying functional dependence. For the static wind velocity Thomson scattering prescription of Section \ref{sec:static_model}, we used the mass-loss rate as a proxy for the star's brightness. The simplest assumption is using precisely the same functional form for the luminosity. Thus, we take
\begin{equation}
    L_*(t)=2\cdot\frac{1+(\eta-1)\sin^2(\frac{\omega}{2} t)}{\eta+1}\bar{L}_*.
\end{equation}
Here, $\bar{L}_*$ is the average luminosity of the star. Also, $n_\text{dust}\sigma_\text{scat}$ is the product of the number density of dust and average scattering cross-section. This gives an average scattering cross-section per unit volume. We make two more approximate assumptions; we take the amount of dust to be a fixed fraction of the amount of gas, $n_\text{dust} = f_\text{d/g} n $, and the average scattering cross-section is assumed to be averaged over all relevant quantities and is therefore constant.

\begin{figure}[t!]
    \centering
    \includegraphics[width=\columnwidth]{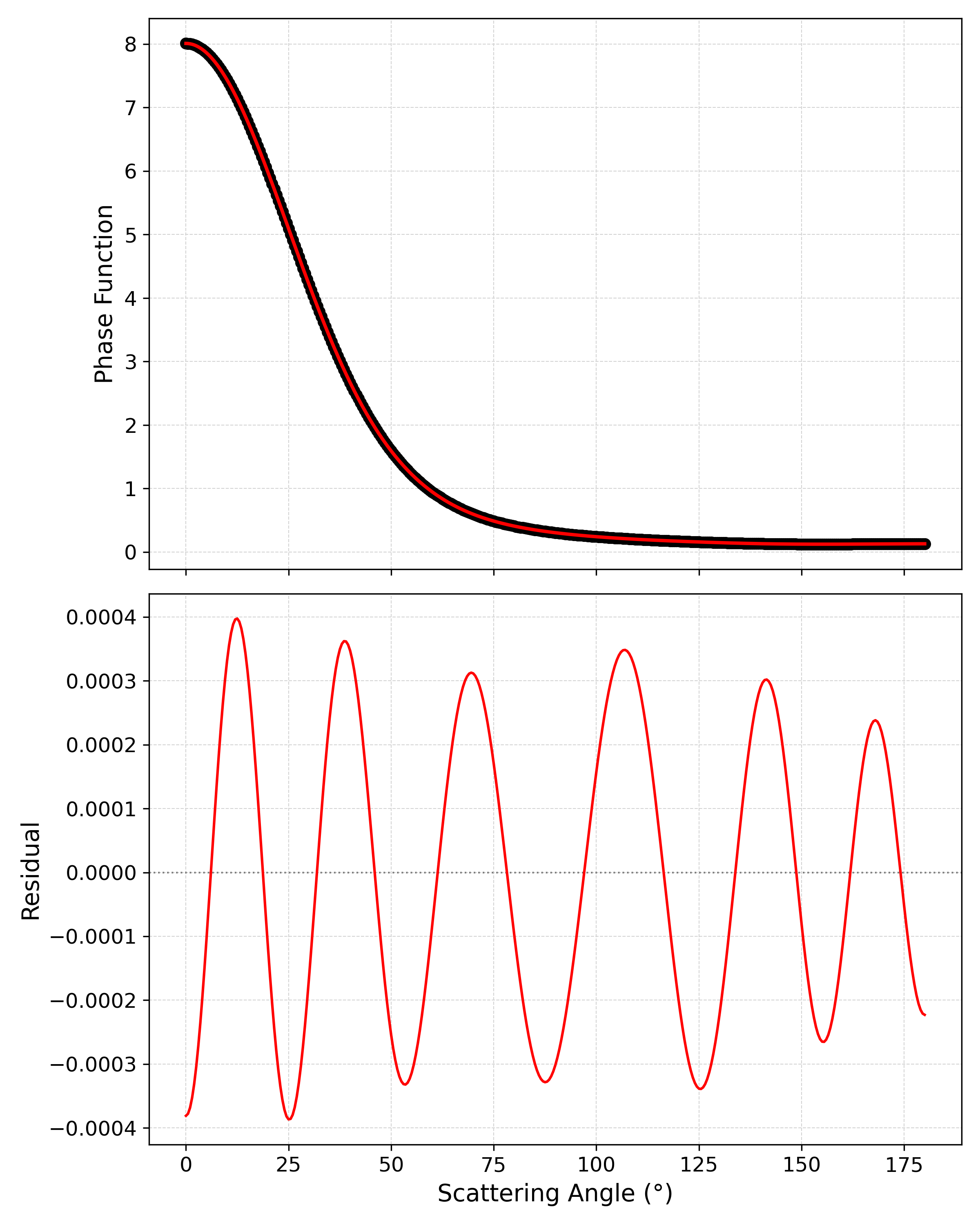}
    \caption{The phase function, $\Phi(\chi)$, determined using the code \textsc{PyMieScatt}. Data from the code are black circles, and our fit using the Taylor series in equation \eqref{Phase_function} for $N=25$ is given by the solid red line. The residuals are also plotted as a function of the scattering angle to show the accuracy of the fit.}
    \label{fig:Phase_Fit}
\end{figure}

\begin{figure}[t!]
    \centering
    \includegraphics[width=\columnwidth]{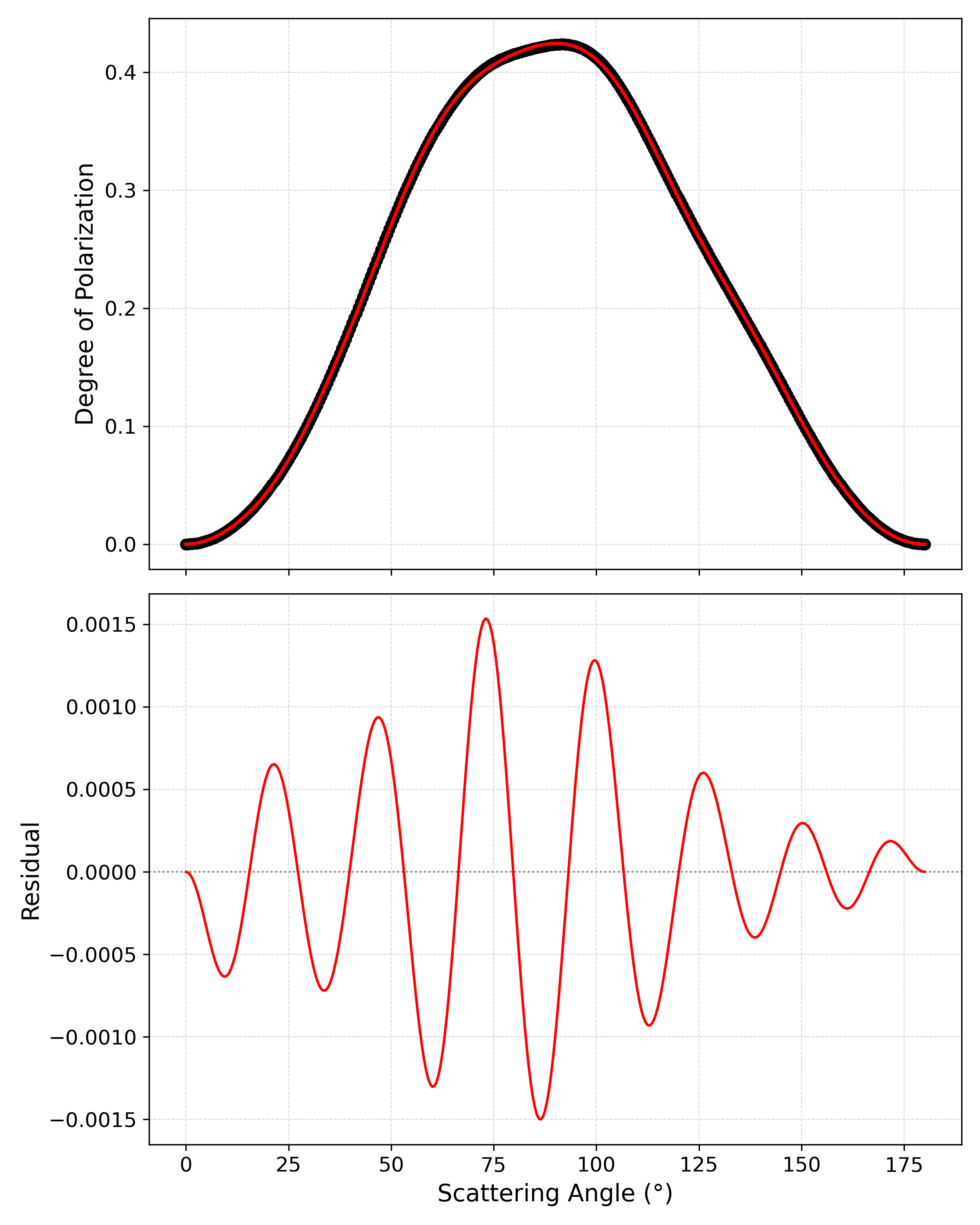}
    \caption{The degree of polarization, $p(\chi)$, determined using the code \textsc{PyMieScatt}. Data from the code are black circles, and our fit using the Taylor series in equation \eqref{Polarization_fraction} for $N=25$ is given by the solid red line. The residuals are also plotted as a function of the scattering angle to show the accuracy of the fit.}
    \label{fig:DoP_Fit}
\end{figure}

We simplify further by assuming that dust from V CVn is amorphous carbon, with an average refractive index of approximately $m_\text{ref}=1.85 +0.5i$ over visible wavelengths \citep{1991ApJ...377..526R}. We use the software \textsc{PyMieScatt} \citep{SUMLIN2018127} to calculate the phase function for the scattering, $\Phi(\chi)$, and the polarization fraction, $p(\chi)$, where $\chi$ is the scattering angle. For simplicity, in \textsc{PyMieScatt}, we averaged over a uniform grain distribution of sizes from $50\,\text{nm}$ to $500\,\text{nm}$, and computed a weighted average from a black body spectrum of $T=3200\,\text{K}$ for V CVn at each grain size for visible wavelengths. We fit the phase function and polarization fraction results from \textsc{PyMieScatt} with a Taylor series of order $N$ in $\cos(\chi)$ as follows,
\begin{equation}\label{Phase_function}
    \Phi(\chi) = \sum_{n=0}^N a_n\cos^n(\chi),
\end{equation}
\begin{equation}\label{Polarization_fraction}
    p(\chi) = (1-\cos^2(\chi))\sum_{m=0}^Nb_m\cos^m(\chi).
\end{equation}
The polarization fraction's expansion is augmented with a clamping term, $(1-\cos^2(\chi))$, to ensure the conditions for the polarization fraction at $\chi=0$ and $\chi=\pi$ are met. Figure \ref{fig:Phase_Fit} and \ref{fig:DoP_Fit} are the results of \textsc{PyMieScatt} and equations \ref{Phase_function} and \ref{Polarization_fraction}, respectively. The coefficients $a_n$ and $b_n$ are listed in table \ref{tab:taylor-coeffs}.

\begin{table}
\centering
\caption{Taylor Series coefficients $a_n$ and $b_n$, for $N=25$.}
\label{tab:taylor-coeffs}
\begin{tabular}{ccc}
\hline
\multicolumn{1}{c}{$n$} & \multicolumn{1}{c}{$a_n$} & \multicolumn{1}{c}{$b_n$} \\
\hline
0  & 0.30257   & 0.42431   \\
1  & 0.45117   & -0.00336  \\
2  & 0.64328   & 0.05888   \\
3  & 0.92121   & 0.63210   \\
4  & 1.19682   & -0.29327  \\
5  & 1.29256   & -1.46474  \\
6  & 1.03804   & 0.41695   \\
7  & 0.82983   & 0.34694   \\
8  & 0.63511   & 0.02292   \\
9  & 0.38037   & 0.69536   \\
10 & 0.29344   & -0.18639  \\
11 & 0.11295   & 0.33456   \\
12 & 0.07694   & -0.16745  \\
13 & -0.00557  & -0.07066  \\
14 & -0.03258  & -0.06901  \\
15 & -0.03599  & -0.30156  \\
16 & -0.06866  & 0.01920   \\
17 & -0.02556  & -0.34520  \\
18 & -0.06026  & 0.06634   \\
19 & -0.00414  & -0.25049  \\
20 & -0.02818  & 0.07123   \\
21 & 0.01154   & -0.07179  \\
22 & 0.01391   & 0.04338   \\
23 & 0.01330   & 0.14786   \\
24 & 0.05758   & -0.00596  \\
25 & -0.00193  & 0.37886   \\
\hline
\end{tabular}
\end{table}

To compute the scattering angle, $\chi$, in this geometry, we assume that the direction to Earth is along the x-axis. This is roughly true for V CVn, as its velocity is almost entirely tangential to the plane of the sky. Thus, the scattering angle, which is the angle between the vector from the source to the scattering event and the scattering event to its final destination, will be formed by the dot product between these vectors,
\begin{equation}
    \cos(\chi) = \hat{\bm{r}}\cdot\hat{\bm{x}}=\sin(\theta)\cos(\phi).
\end{equation}
Now, we may form a complete differential scattering contribution for each cell in the domain,
\begin{align}\label{dI}
    dI_\text{scat}&=\frac{f_\text{d/g}\braket{\sigma_\text{scat}}}{4\pi\bar{m}}\cdot L_*(t-\tfrac{r}{c})\rho(r,\theta)
    \nonumber
    \\&\times\sum_{n=0}^N a_n\sin^n(\theta)\cos^n(\phi) \,d\phi \,d\theta\, dr
\end{align}
Of this light, some of what is scattered will become polarized. The quantity $p(\chi)$ is the fraction of scattered light which is linearly polarized. Now, we can define the differential fraction of light which is polarized, 
\begin{equation}\label{dI_pol}
    dI_\text{pol}=p(\chi)dI_\text{scat}.
\end{equation}
With $dI_\text{pol}$, we define differential contributions of the Stokes parameters, $Q$ and $U$ from each cell in the computational domain. In a typical way for linear polarization (see, e.g. \cite{Chandrasekhar1960}), these are
\begin{equation}\label{dQ}
    dQ = p(\chi)\cos(2\psi)dI_\text{scat},
\end{equation}
\begin{equation}\label{dU}
    dU = p(\chi)\sin(2\psi)dI_\text{scat}.
\end{equation}
Here, $\psi$ is the typical definition of the polarization angle on the sky, so it vanishes at $+z$ and goes counterclockwise. The scattering plane is spanned by the vectors $\bm{r}$ and $\bm{x}$. The electric field will be oriented perpendicular to this plane along the normal vector,

\begin{equation}
    \hat{\bm{n}}=\hat{\bm{x}}\times\hat{\bm{r}}=\sin(\theta)\sin(\phi)\hat{\bm{z}}-\cos(\theta)\hat{\bm{y}}.
\end{equation}
Conveniently, this is in the plane of the sky. Using the components of $\hat{\bm{n}}$, some trigonometric identities, and algebra gives
\begin{align}\label{cos_psi}
    \cos(2\psi) &=\cos^2(\psi)-\sin^2(\psi), \nonumber
    \\ &= \frac{\sin^2(\theta)\sin^2(\phi)-\cos^2(\theta)}{\sin^2(\theta)\sin^2(\phi)+\cos^2(\theta)},
\end{align}
\begin{align}\label{sin_psi}
    \sin(2\psi) &=2\sin(\psi)\cos(\psi), \nonumber
    \\& =-\frac{2\sin(\theta)\cos(\theta)\sin(\phi)}{\sin^2(\theta)\sin^2(\phi)+\cos^2(\theta)}.
\end{align}
Using equations \eqref{Phase_function}, \eqref{Polarization_fraction}, \eqref{dI}, \eqref{dI_pol}, \eqref{dQ}, \eqref{dU}, \eqref{cos_psi}, and \eqref{sin_psi}, we find
\begin{align}
    I(t) &= \frac{f_\text{d/g}\braket{\sigma_\text{scat}}}{4\pi\bar{m}} \sum_{n=0}^N a_n\int_0^\infty dr\, L_*(t-\tfrac{r}{c})
    \nonumber
    \\& \times\int_0^\pi d\theta\,\rho(r,\theta)\sin^n(\theta)
    \int_0^{2\pi} d\phi\,
    \cos^n(\phi) ,
\end{align}
\begin{align}
    Q(t) &= \frac{f_\text{d/g}\braket{\sigma_\text{scat}}}{4\pi\bar{m}} \sum_{n=0}^N \sum_{m=0}^N a_nb_m\int_0^\infty dr\, L_*(t-\tfrac{r}{c})
    \nonumber
    \\& \times \int_0^\pi d\theta\, \rho(r,\theta)\sin^{m+n}(\theta)
    \nonumber
    \\& \times \int_0^{2\pi} d\phi\,\biggl\{
    \cos^{m+n}(\phi)[1-\sin^2(\theta)\cos^2(\phi)]
    \nonumber
    \\&
    \times\frac{[\sin^2(\theta)\sin^2(\phi)-\cos^2(\theta)]}{\sin^2(\theta)\sin^2(\phi)+\cos^2(\theta)} \biggr\},
\end{align}
\begin{align}
    U(t) &= -\frac{f_\text{d/g}\braket{\sigma_\text{scat}}}{2\pi\bar{m}} \sum_{n=0}^N \sum_{m=0}^N a_nb_m\int_0^\infty dr\, L_*(t-\tfrac{r}{c})
    \nonumber
    \\& \times\int_0^\pi d\theta\, \rho(r,\theta)\sin^{m+n+1}(\theta)\cos(\theta)
    \\& \times\int_0^{2\pi} d\phi\,
    \frac{\cos^{m+n}(\phi)[1-\sin^2(\theta)\cos^2(\phi)]\sin(\phi)}{\sin^2(\theta)\sin^2(\phi)+\cos^2(\theta)}.\nonumber
\end{align}
Utilizing the power formulae for cosine \citep{Beyer1987} and some algebra, the azimuthal integral for $I$ reduces significantly, with only even terms contributing. The expression becomes
\begin{align}\label{eq:I}
    I(t) &= \frac{f_\text{d/g}\braket{\sigma_\text{scat}}}{2\bar{m}} \sum_{n=0}^{\lfloor \tfrac{N}{2} \rfloor} \frac{{(2n)!}\,a_{2n}}{2^{2n}(n!)^2}\int_0^\infty dr\, L_*(t-\tfrac{r}{c})
    \nonumber
    \\& \times\int_0^\pi d\theta\, \rho(r,\theta)\sin^{2n}(\theta).
\end{align}
The azimuthal integrals for $Q(t)$ and $U(t)$ are also computable. We examine the integrand for the azimuthal integral for $Q(t)$,
\begin{align}
    f_{Q}(\theta,\phi)&=
    \cos^{m+n}(\phi)[1-\sin^2(\theta)\cos^2(\phi)]
    \\&
    \nonumber
    \times\frac{\sin^2(\theta)\sin^2(\phi)-\cos^2(\theta)}{\sin^2(\theta)\sin^2(\phi)+\cos^2(\theta)}.
\end{align}
First, we note that $f_{Q}(\theta,\phi-\pi)=f_Q(\theta,\pi-\phi) \,\forall\phi$, which means that the integral over $\phi \in[0,2\pi]$ is twice the integral over $\phi \in[0,\pi]$. Next, we note that $f_{Q}(\theta,\phi-\tfrac{\pi}{2})=(-1)^{m+n}f_Q(\theta,\tfrac{\pi}{2}-\phi)\, \forall\phi$, which means that the integral vanishes for all odd combinations of $m+n$. Therefore, only even combinations of $m+n$ contribute, and these are symmetric, so the integral over $\phi \in[0,\pi]$ is twice the integral over $\phi \in[0,\tfrac{\pi}{2}]$. Thus, 
\begin{align}
    Q(t) &= \frac{f_\text{d/g}\braket{\sigma_\text{scat}}}{\pi\bar{m}} \sum_{\substack{n=0, m=0 \\ m + n = 2k}}^N a_nb_m\int_0^\infty dr\, L_*(t-\tfrac{r}{c})
    \nonumber
    \\& \times\int_0^\pi d\theta\, \rho(r,\theta)\sin^{m+n}(\theta) \int_0^{\tfrac{\pi}{2}} d\phi\, f_Q(\theta,\phi).
\end{align}
After some careful algebra and the integral tables of \cite{GradshteynRyzhik2014}, we get
\begin{align}\label{eq:Q}
    Q(t) &= \frac{f_\text{d/g}\braket{\sigma_\text{scat}}}{2\pi\bar{m}} \sum_{\substack{n=0, m=0 \\ m + n = 2k}}^N a_nb_m\int_0^\infty dr\, L_*(t-\tfrac{r}{c})
    \nonumber
    \\& \times\int_0^\pi d\theta\, \rho(r,\theta)\sin^{m+n}(\theta) \Xi(\theta;m,n),
\end{align}
where
\begin{align}
    &\Xi(\theta;m,n) 
    \nonumber
    \\&= s^2_\theta B(\tfrac{3}{2},\tfrac{m+n+1}{2}){_2}F_1[\tfrac{m+n+1}{2},1;\tfrac{m+n+4}{2};s^2_\theta] 
    \nonumber
    \\& -c^2_\theta B(\tfrac{1}{2},\tfrac{m+n+1}{2}){_2}F_1[\tfrac{m+n+1}{2},1;\tfrac{m+n+2}{2};s^2_\theta] 
    \nonumber
    \\& -s^4_\theta B(\tfrac{3}{2},\tfrac{m+n+3}{2}){_2}F_1[\tfrac{m+n+3}{2},1;\tfrac{m+n+6}{2};s^2_\theta]
    \nonumber
    \\& +s^2_\theta c^2_\theta B(\tfrac{1}{2},\tfrac{m+n+3}{2}){_2}F_1[\tfrac{m+n+3}{2},1;\tfrac{m+n+4}{2};s^2_\theta].
\end{align}
Here, $s_\theta$ and $c_\theta$ are short-hand notation for $\sin(\theta)$ and $\cos(\theta)$ respectively, $B(k_1,k_2)$ is Euler's beta function, and ${_2}F_1(k_3,k_4;k_5;\xi)$ is Gauss' hypergeometric function (see, e.g. \cite{arfken2012mathematical}). Despite the complicated form of $\Xi(\theta;m,n)$, it is significantly faster to compute, compared to numerically integrating $f_Q(\theta,\phi)$ over $\phi$, for each value of $\theta$.

Thankfully, $U(t)$ turns out to be far simpler to compute due to the functional form of the integrand for the azimuthal integral,
\begin{equation}
    f_{U}(\theta,\phi)=
    \frac{\cos^{m+n}(\phi)[1-\sin^2(\theta)\cos^2(\phi)]\sin(\phi)}{\sin^2(\theta)\sin^2(\phi)+\cos^2(\theta)}.
\end{equation}
We note that $f_U(\theta,\phi-\pi)=-f_U(\theta,\pi-\phi)\,\forall\phi$. Therefore, the integral over $\phi \in[0,2\pi]$ vanishes. Thus, under the azimuthal symmetric density field,
\begin{equation}\label{eq:U}
    U(t) = 0.
\end{equation}
Finally, we may form the observed polarization signal in the usual way, namely, 
\begin{equation}\label{Mie_Observed_Polarization}
    P_R=\frac{\sqrt{Q^2(t)+U^2(t)}}{I(t)+I_*(t)}=\frac{|Q(t)|}{I(t)+I_*(t)}.
\end{equation}
Here, we've added the star's unpolarized intensity, $I_*$, to the scattered intensity, $I(t)$. From our model in \textsc{PyMieScatt}, we also calculate the averaged Mie scattering cross section, $\braket{\sigma_\text{scat}}=9.23\times10^{-10}\,\text{cm}^2$. We also take an approximate dust-to-gas ratio on the order of $f_\text{d/g}\approx10^{-3}$ (see, e.g. \citet{Winters2000, Hofner2018}).


\section{Simulation Results and Discussion}\label{sec:Results_and_Discussion}
\begin{figure}[t!]
    \centering
    \includegraphics[width=\columnwidth]{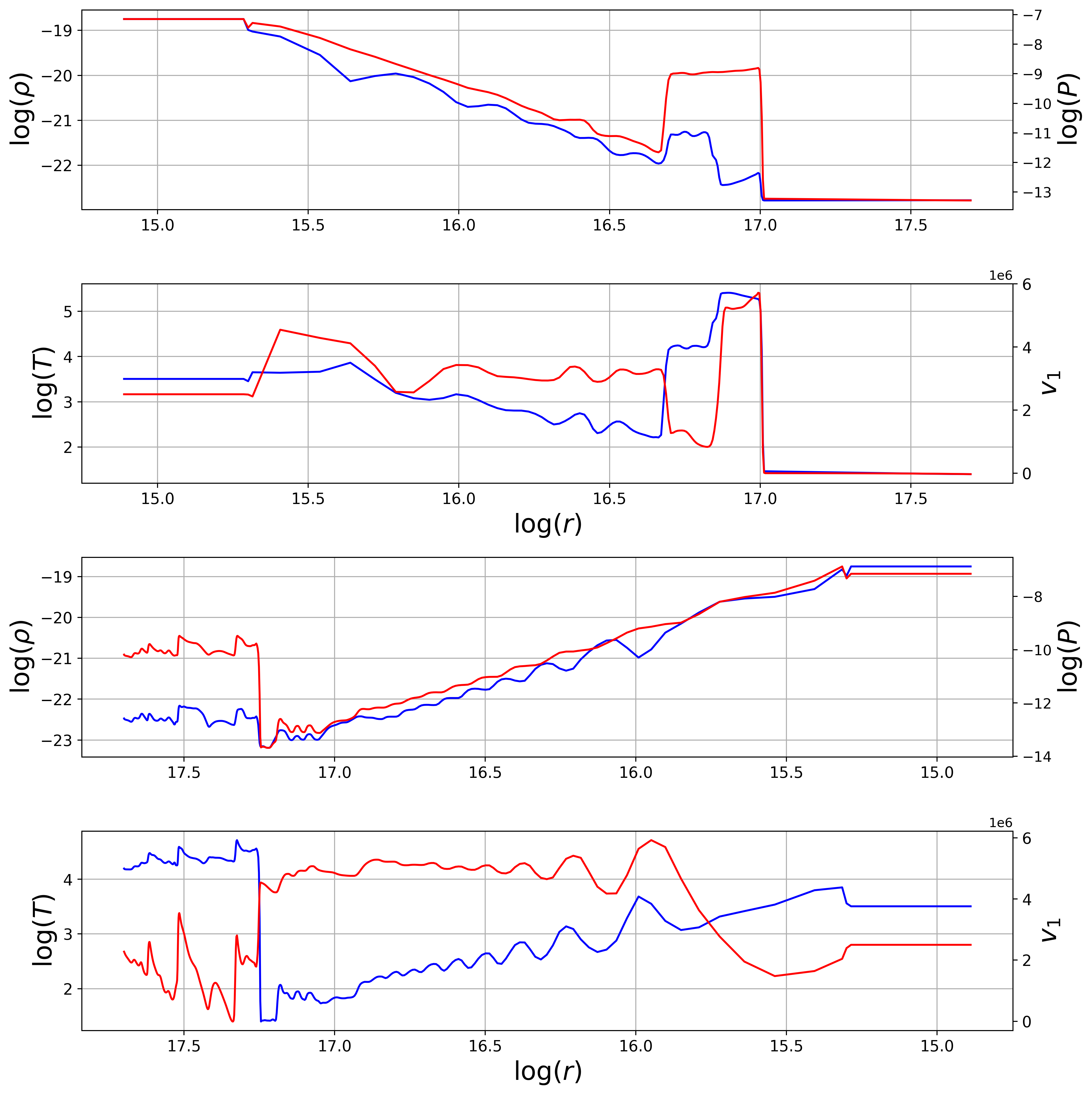}
    \caption{From top to bottom, the 1-dimensional slices along $\theta=\pi/2$ (perpendicular to the direction of motion of V CVn) of the logarithms of the density (blue), thermal pressure (red), temperature (blue), and radial velocity (red) at time $t=4.3750\times10^{11}\,\text{s}$. All are plotted against the logarithm of the radial coordinate. Once again, from top to bottom for the lower two plots, the 1-dimensional slices along $\theta=\pi$ (opposed to the direction of motion of V CVn) of the logarithms of the density blue, thermal pressure red, temperature blue, and radial velocity red at time $t=4.3750\times10^{11}\,\text{s}$. All are plotted against the logarithm of the radial coordinate, but the axis is reversed, so it is easier to imagine going from the star through the tail. All units are cgs unless otherwise indicated.}
    \label{9}
\end{figure}

\begin{figure}[t!]
    \centering
    \includegraphics[width=\columnwidth]{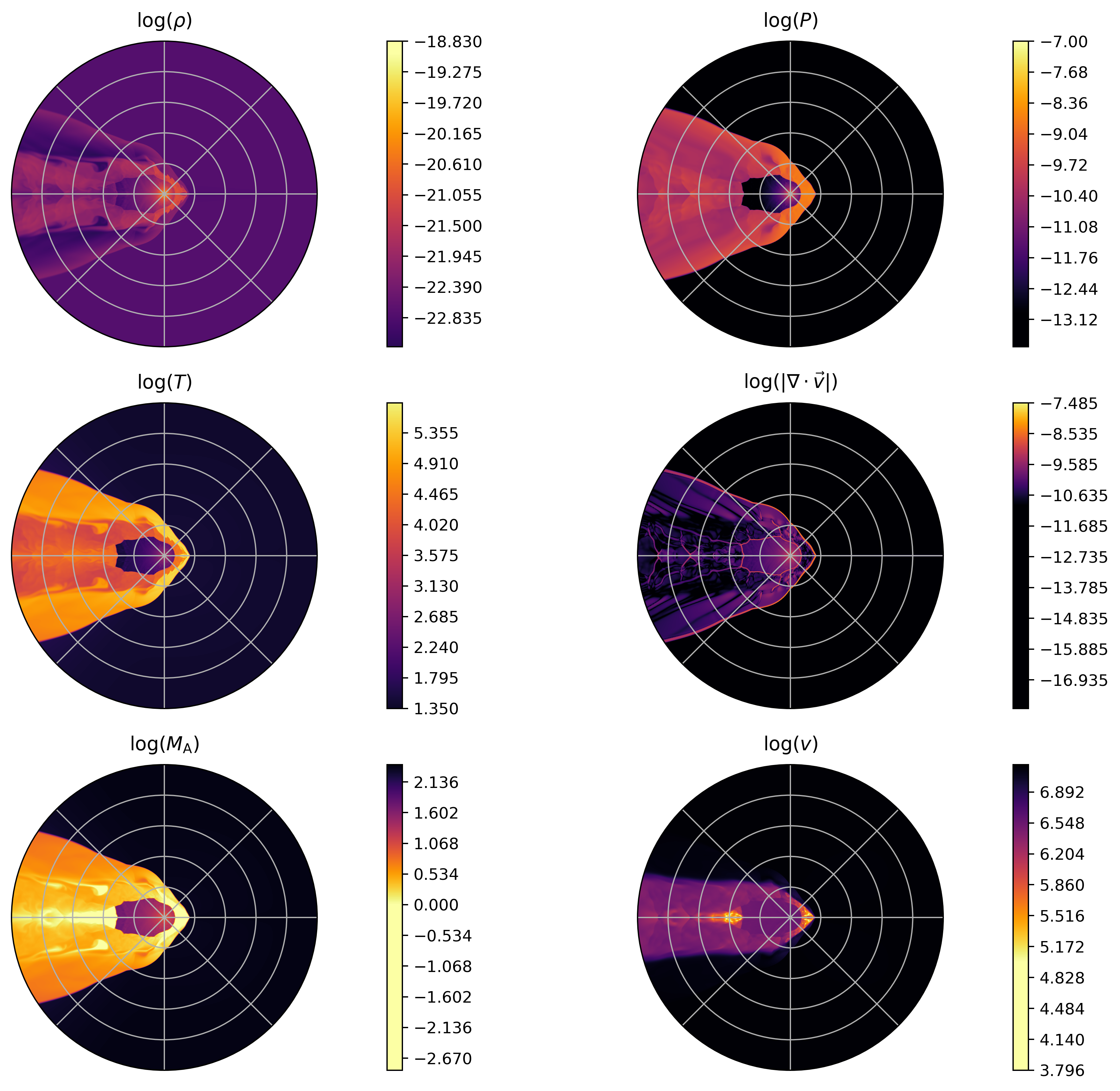}
    \caption{From left to right, top to bottom, this Figure represents a heat map of the logarithms of density, thermal pressure, temperature, absolute velocity divergence, Mach number, and speed, at a problem time of $t=7.0000\times10^{11}\,\text{s}$ for the static wind velocity case. Each concentric circle (grey) indicates a distance of $1.0\times10^{17}\,\text{cm}$, and each radial line (grey) indicates an angular increment of $\pi/4$ radians. All units are cgs unless otherwise indicated.}
    \label{10}
\end{figure}

\begin{figure}[t!]
    \centering
    \includegraphics[width=\columnwidth]{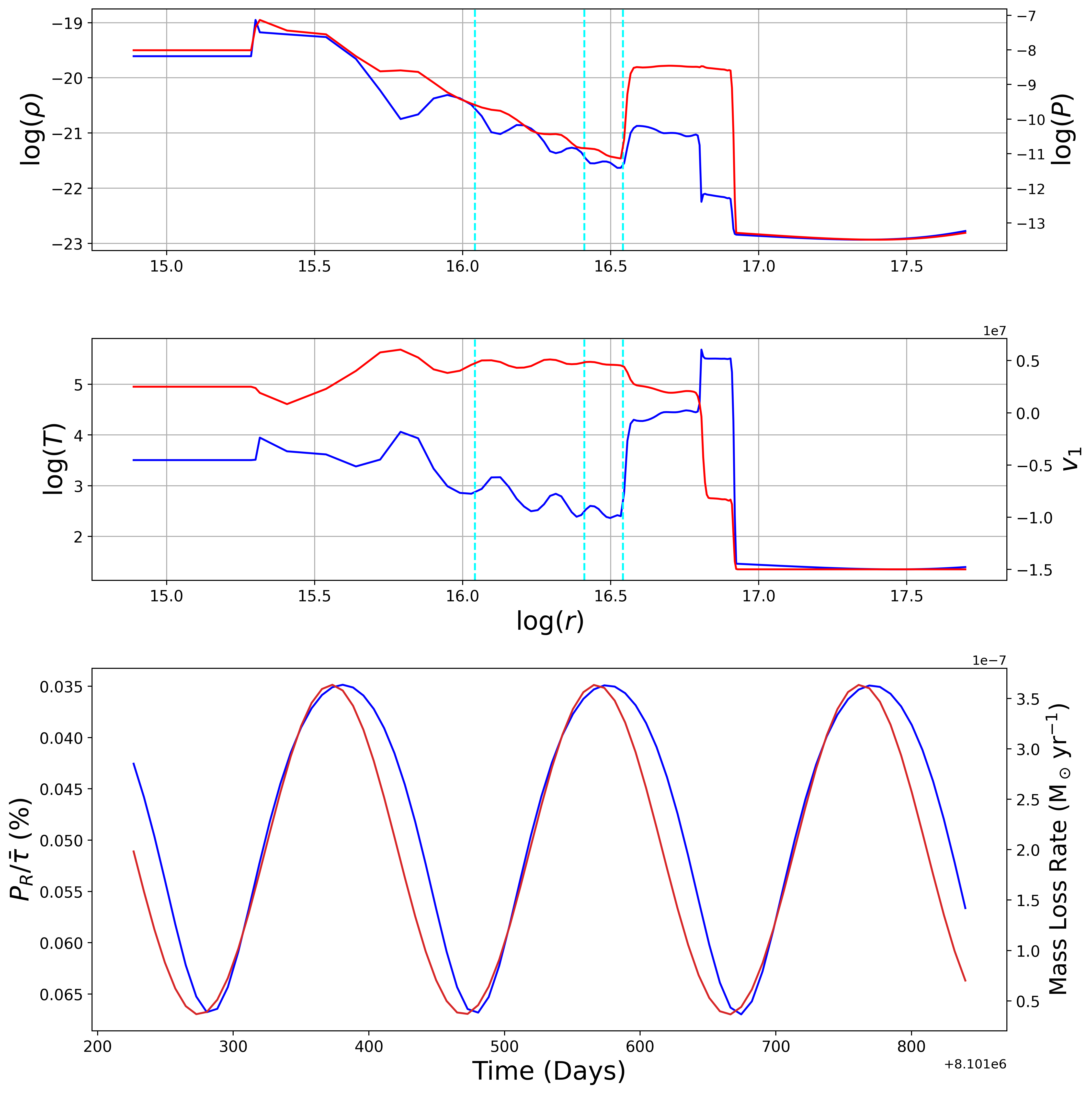}
    \caption{From top to bottom, the 1-dimensional slices along $\theta=0$ (the direction of motion of V CVn) of the logarithms of the density blue, thermal pressure red, temperature blue, and radial velocity red at time $t=7.0000\times10^{11}\,\text{s}$. All are plotted against the logarithm of the radial coordinate. The bottom plot from left to right is the residual polarization as a fraction of the optical depth in percentage blue and the mass-loss rate of the star red. The time coordinates for this plot are the nearest 80 points to the time of the slice. All units are cgs unless otherwise indicated.}
    \label{11}
\end{figure}

\begin{figure}[t!]
    \centering
    \includegraphics[width=\columnwidth]{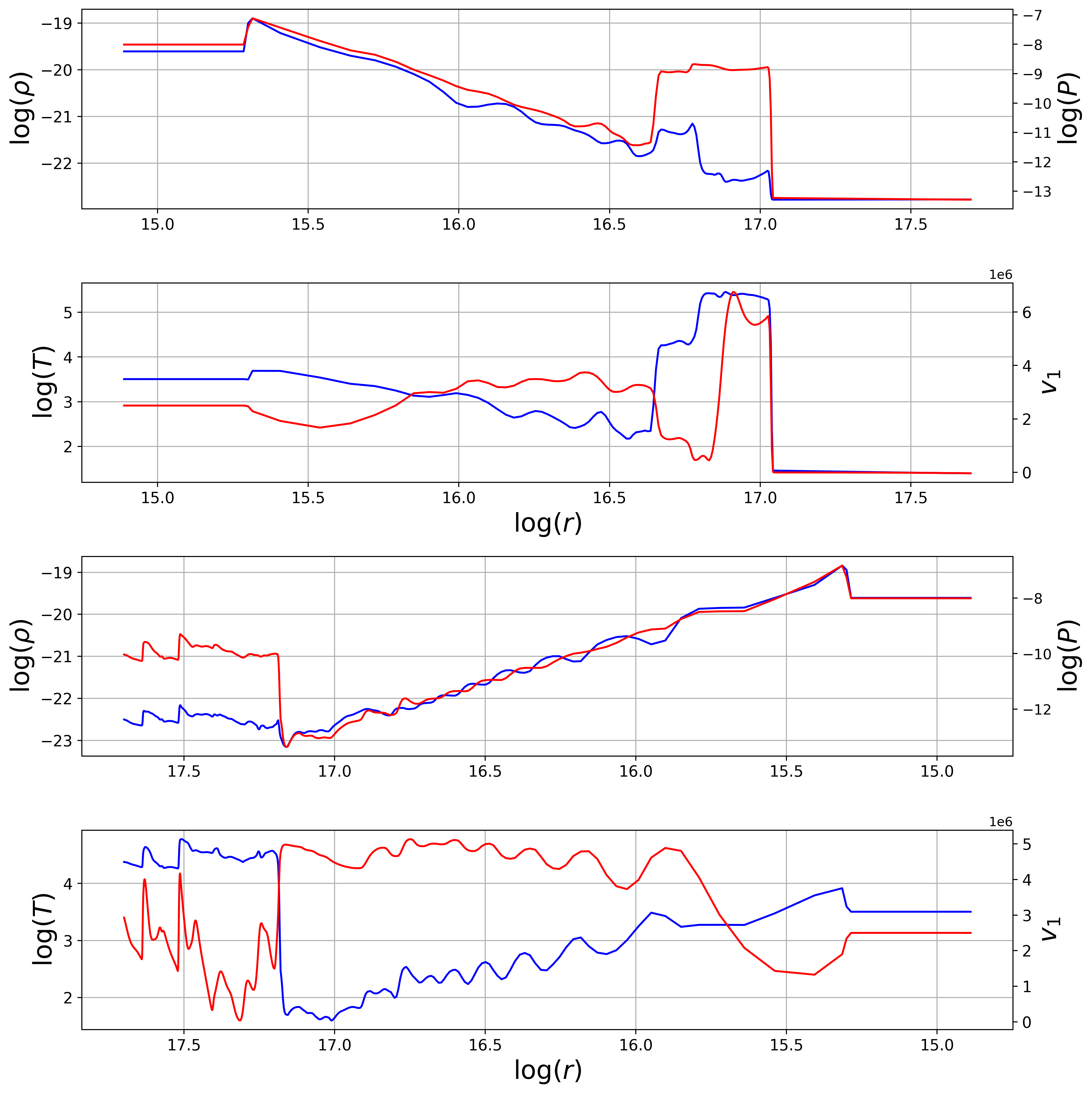}
    \caption{From top to bottom, the 1-dimensional slices along $\theta=\pi/2$ (perpendicular to the direction of motion of V CVn) of the logarithms of the density blue, thermal pressure red, temperature blue, and radial velocity red at time $t=7.0000\times10^{11}\,\text{s}$. All are plotted against the logarithm of the radial coordinate. Once again, from left to right, top to bottom for the lower two plots, the 1-dimensional slices along $\theta=\pi$ (opposed to the direction of motion of V CVn) of the logarithms of the density blue, thermal pressure red, temperature blue, and radial velocity red at time $t=7.0000\times10^{11}\,\text{s}$. All are plotted against the logarithm of the radial coordinate, but the axis is reversed, so it is easier to imagine going from the star through the tail. All units are cgs unless otherwise indicated.}
    \label{12}
\end{figure}

In the simulations, the stellar wind is spherically symmetric and has a mass-loss rate which varies periodically. The star moves directly along the z-axis (from left to right in the simulations). In the rest frame of the star, this implies that the ISM is moving with the star's velocity but in the opposing direction. Therefore, the ISM is moving from right to left in the simulations. 
\begin{figure}[t!]
    \centering
    \includegraphics[width=\columnwidth]{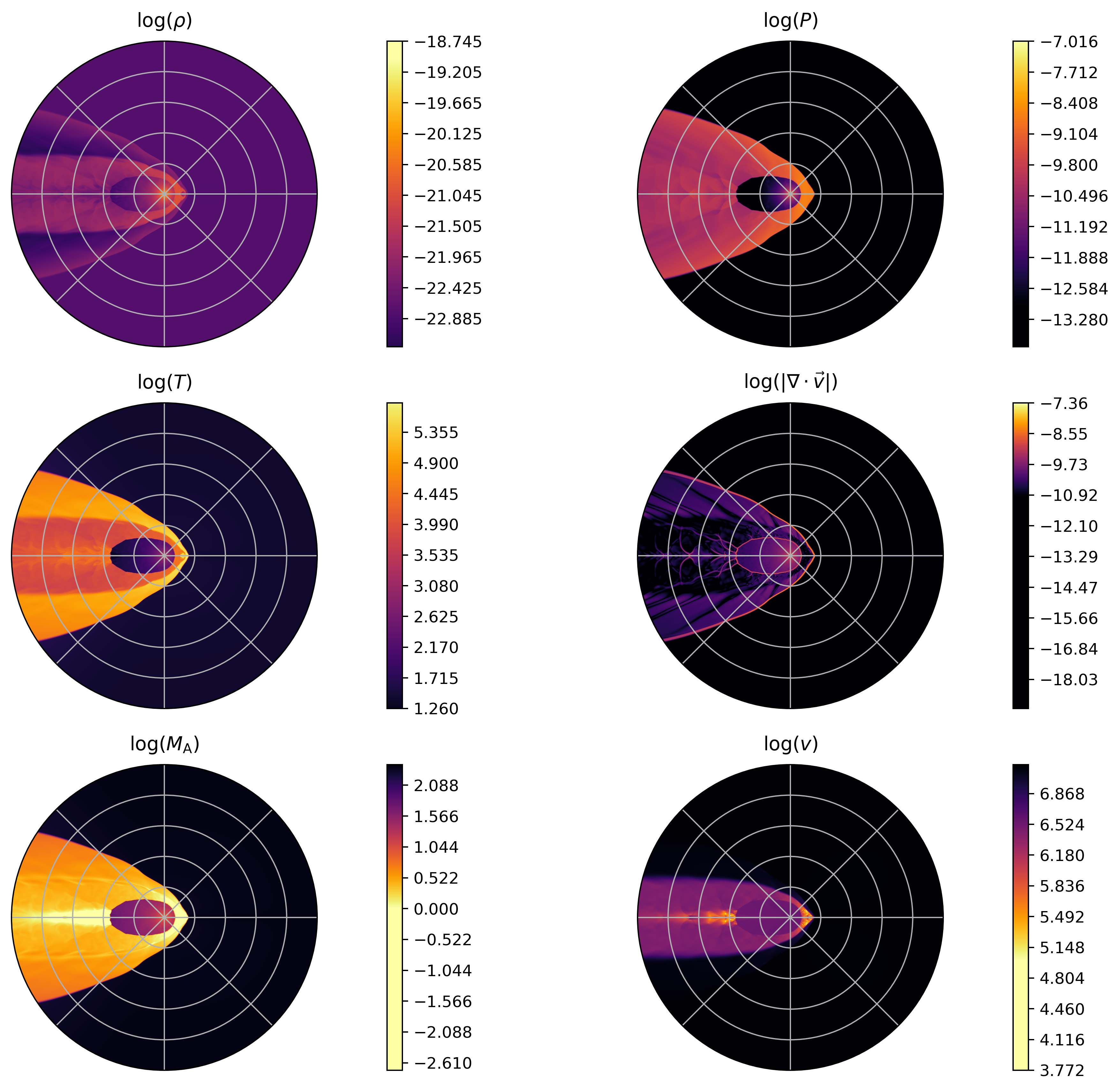}
    \caption{From left to right, top to bottom, this Figure represents a heat map of the logarithms of density, thermal pressure, temperature, absolute velocity divergence, Mach number, and speed, at a problem time of $t=4.3750\times10^{11}\,\text{s}$ for the static wind velocity case. Each concentric circle (grey) indicates a distance of $1.0\times10^{17}\,\text{cm}$, and each radial line (grey) indicates an angular increment of $\pi/4$ radians. All units are cgs unless otherwise indicated.}
    \label{7}
\end{figure}
To understand the dynamics of the interaction, it is useful to make some order of magnitude approximations and define important timescales and dimensionless ratios for the flow using Table \ref{tab:Variable_Params}. We define a characteristic flow time from the star to the bow shock as $t_\text{flow}\approx\bar{R}_\text{SO}/\bar{v}_\text{w}\approx300\,\text{yr}$. The pulsation period of the wind is $\tau\approx0.5\,\text{yr}$. Thus, by forming a dimensionless ratio $\tau / t_\text{flow} \approx 2.0\times10^{-3}$, it becomes clear that the region of free wind expansion from the star to the bow shock evolves through a series of many rapid oscillations. The dimensionless ratio of the ISM speed to the stellar wind speed is $v_*/\bar
{v}_\text{w}=6$. An important consideration is the post-shock cooling time for both the forward and reverse shocks, compared to the time it takes the shocked material to flow across the shocked region to the contact discontinuity. This dimensionless ratio determines whether radiative cooling is important for the forward and reverse shocks. To get an order-of-magnitude estimate for the cooling time of the shocks, we assume that the thickness of the forward and reverse shocks will be some fraction, say, a quarter of the standoff distance to the bow shock, $\bar{R}_\text{SO}/4$. In strong shock conditions for $\gamma=5/3$, the post-shock velocities are about a quarter of the pre-shock velocities. Thus, we may estimate that the flow speed for the reverse shock is $\bar{v}_w/4$, and the flow speed for the forward shock is $v_*/4$. Therefore, the flow time scale for the reverse shock as $t_\text{flow,r}\approx(\bar{R}_\text{SO}/4)/(\bar{v}_\text{w}/4)\approx 300\,\text{yr}$, and the flow time scale for the forward shock is $t_\text{flow,f}\approx(\bar{R}_\text{SO}/4)/(\bar{v}_*/4)\approx 50\,\text{yr}$; meaning the flow time for the reverse shock is on the order of hundreds of years, while the flow time for the forward shock is on the order of tens of years in an ideal case. For the cooling time estimates of the forward and reverse shocks, we turn to \cite{Draine_Book}, which for the reverse shock of $\bar{v}_\text{w}\approx30\,\text{km}\,\text{s}^{-1}$, an estimate of the cooling time is $t_\text{cool,r}\approx10^{4}[\text{cm}^{-3}/n_s]\,\text{yr}$. Here, $n_s$ is the pre-shock number density, which we may estimate as $\dot{\bar{M}}/4\pi \bar{m}R_\text{SO}^2v_\text{w}$, since in this regime, the wind is free to expand. Thus, we estimate the cooling time for the reverse shock as $t_\text{cool,r}\approx 15\,\text{yr}$, meaning the reverse shock cools on the order of tens of years. For the forward shock, which has a much higher speed \cite{Draine_Book} provides a simple estimate, $t_\text{cool,f}\approx7000(\text{cm}^{-3}/n_0)(v_*/100\,\text{km}\,\text{s}^{-1})^{3.4}\,\text{yr}\approx2800\,\text{yr}$, meaning the forward shock cools on the order of thousands of years. Forming dimensionless ratios of approximate flow time across the shocks to the cooling times of the shocks, we get $t_\text{flow,r}/t_\text{cool,r}\approx20$ and $t_\text{flow,f}/t_\text{cool,f}\approx2\times10^{-2}$. Therefore, the forward shock is essentially adiabatic and has no time to cool, whereas the reverse shock is extremely radiative. Though we do not take radiative cooling into account in our simulations, we note that the reverse shock would cool very quickly and, though it would be a much thinner region, it would condense far beyond the strong shock density ratio limit of four, possibly contributing more to the polarization signal. Moreover, radiative instabilities could increase the envelope's asymmetry, contributing to the polarization signal. 
\begin{figure}[t!]
    \centering
    \includegraphics[width=\columnwidth]{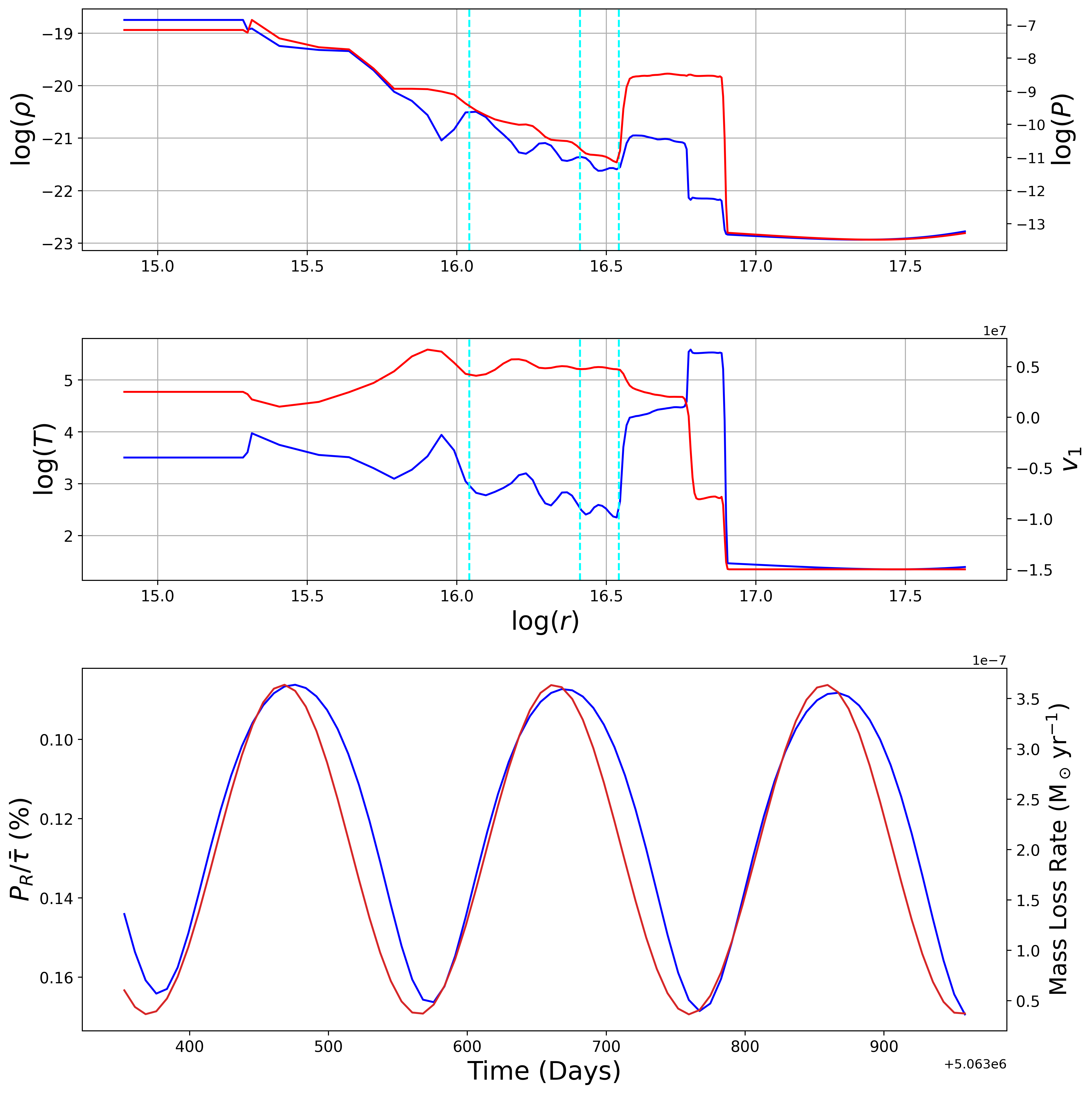}
    \caption{From top to bottom, the 1-dimensional slices along $\theta=0$ (the direction of motion of V CVn) of the logarithms of the density (blue), thermal pressure (red), temperature (blue), and radial velocity (red) at time $t=4.3750\times10^{11}\,\text{s}$. All are plotted against the logarithm of the radial coordinate. The bottom plot is the residual polarization as a fraction of the optical depth as a percentage (blue) and the mass-loss rate of the star (red). The time coordinates for this plot are the nearest 80 points to the time of the slice. All units are cgs unless otherwise indicated.}
    \label{8}
\end{figure}
Figures \ref{7} through \ref{12} represent the static wind velocity case.\footnote{For an example of time evolution for the density field from V CVn, we refer the reader to the supplementary material for an animation of the density.} The interaction between the wind from V CVn and the oncoming ISM produces an incredibly rich structure and physics, far more complicated than the analytic bow shock model of \cite{Wilkin1996} would indicate. By looking at the $\theta = 0$ slice plots (Figures \ref{8}, \ref{11}), one may discern the different regions of the interaction.
Moving from left to right in the $\theta=0$ slice plots, the flat line for each variable represents the prescribed values of the variables in the wind bubble boundary. Immediately following the wind bubble boundary zone is a region where the flow tends to get less dense and cool, and thermal pressure decreases while the velocity increases. This is a rarefaction fan. In this region, the supersonic wind can spread out and rarefy. The oscillations in the rarefaction fan are due to the oscillations in the boundary conditions. In the case of a stellar wind that is constant in time, one would likely see a smooth, non-oscillatory curve connecting the wind bubble region with the next region of the interaction. The time variations in the boundary conditions perturb this structure, and the perturbations continue through the rarefaction fan. The next region of the interaction begins at the discontinuity, which occurs in all of the variables. This discontinuity is the reverse shock. Here, the material being driven by the stellar wind is compressed, which vastly increases its density and thermal pressure, heated, which causes the temperature to change by orders of magnitude, and decelerated to a locally sub-sonic speed. The region to the right of the reverse shock is where the heated and compressed stellar wind material gathers.
Further to the right of the reverse shock is a thin region most easily noticed as a discontinuity in density. Though the density is discontinuous, the thermal pressure is not. This indicates a contact discontinuity, which separates the material driven by the stellar wind from the material coming from the ISM. No material may be transferred across a contact discontinuity, as they have the property that the velocity normal to the contact surface vanishes. One may notice by looking at the 1-velocity that this is indeed the case since for the $\theta=0$ slices, the 1-direction is the normal direction. Following to the right of the contact discontinuity is a highly compressed, hot region of ISM gas. This region comprises ISM material swept up by the forward shock, which is the discontinuity in all variables to the right. Once again, one may notice that from the ISM (to the right of the forward shock), the shock sweeps up this material, heating and compressing it, which changes its density, thermal pressure, and temperature by vast amounts. Also, locally, the region between the contact and forward shock is subsonic. Of course, to the right of the forward shock is the yet-to-be-disturbed ISM.      
The heat maps show the density, thermal pressure, temperature, velocity divergence, Mach number, and flow speed from left to right and top to bottom, respectively. Of particular note are the plots of velocity divergence and Mach number. The velocity divergence is a good measure of the location of shocks since they cause rapid compression of the material. Anywhere a sudden compression of material exists, the velocity divergence will take on a large (relatively speaking) negative value. Thus, regions in the heat map of velocity divergence with the highest absolute value will typically represent the shock structure of the interaction. The Mach number, defined as the ratio of the local gas speed to sound speed, showcases sub-sonic and super-sonic flow regions. Since the heat map is logarithmic, zero represents the local sonic transition. Therefore, in areas with a logarithmic Mach number less than zero, the flow is sub-sonic, while in areas with a logarithmic Mach number greater than zero, the flow is super-sonic. One may notice that sub-sonic flow regions typically occur with a high absolute velocity divergence. This is the case, as the shocks normally cause the local sound speed to rise drastically while decelerating the flow, thereby causing the Mach number to decrease. 
Observing the heat maps, particularly the density, thermal pressure, and velocity divergence, one may notice an extremely interesting dynamic structure in the left half of the simulation, behind the star, with respect to its direction of motion. The velocity divergence and thermal pressure plots show that a rarefied region of extremely low thermal pressure develops around the star, surrounded by an asymmetric shock structure which connects to the reverse shock in the front of the star. As time progresses, the star forms a comet-like tail, as one may see by examining the density heat maps. Even more interesting in the region behind the star is the presence of weaker shocks and a cascade of even weaker shocks that change dynamically as the simulation evolves. One may see this structure by following the evolution of the velocity divergence through time. One may notice that by examining the density and temperature plots, the contact discontinuity that separated the stellar wind from the ISM at the front of the star persists into the tail. The hot swept-up ISM material, which travels through the forward shock, is dynamically transported backwards into the outer envelope of the tail.
\begin{figure}[t!]
    \centering
    \includegraphics[width=\columnwidth]{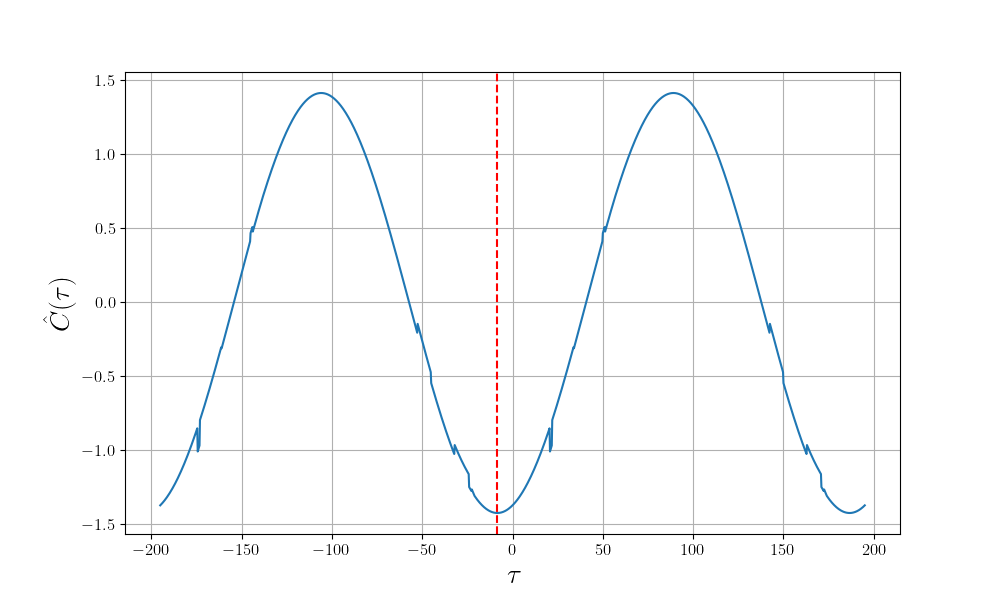}
    \caption{The normalized cross-correlation function of the residual Thomson polarization and mass-loss rate for the static wind case of V CVn blue plotted against the lag time, $\tau_\text{l}$ in days. The dashed line red, located at $\tau_\text{l}\approx -8.5\,\text{days}$ with a corresponding value of $\hat{C}(\tau_\text{l})\approx -1.42274$, represents a local minimum of the function. Due to the normalization condition, this function value means that the residual polarization and mass-loss rate have an extremely strong inverse relationship. Since the local minimum doesn't occur exactly at zero lag, it means that the polarization leads the mass-loss rate by about $8.5$ days. However, they still have an extremely strong inverse relationship across the dataset.}
    \label{Cross_Corr_Static}
\end{figure}
In contrast, the stellar wind material which passes through the reverse shock is transported into the cometary structure. As this material moves through the cometary structure, it then passes through the weaker shocks, but it is still mostly supersonic. Here, a very interesting mixing occurs, which can be most easily noticed in the Mach number, density, and speed plots. The material passing through the oblique shock, still mostly supersonic, forms a shear layer with the subsonic gas, which passes through the shock structure directly behind the star due to the density, temperature, and relative velocity differences at this interface. As time progresses, this interface becomes unstable due to the shear flow, and Kelvin--Helmholtz instabilities \citep{drazin2002introduction} develop and begin to mix the layers. As this mixing occurs, the weaker cascading shocks travelling through this tail region sweep up and mix the Kelvin-Helmholtz unstable regions even more. All of the dynamic features occurring within the tail region of the interaction cause a large amount of turbulence in the flow. Due to its stochastic nature, this turbulence will manifest as an asymmetry in three-dimensional simulations.
\begin{figure}[t!]
    \centering
    \includegraphics[width=\linewidth]{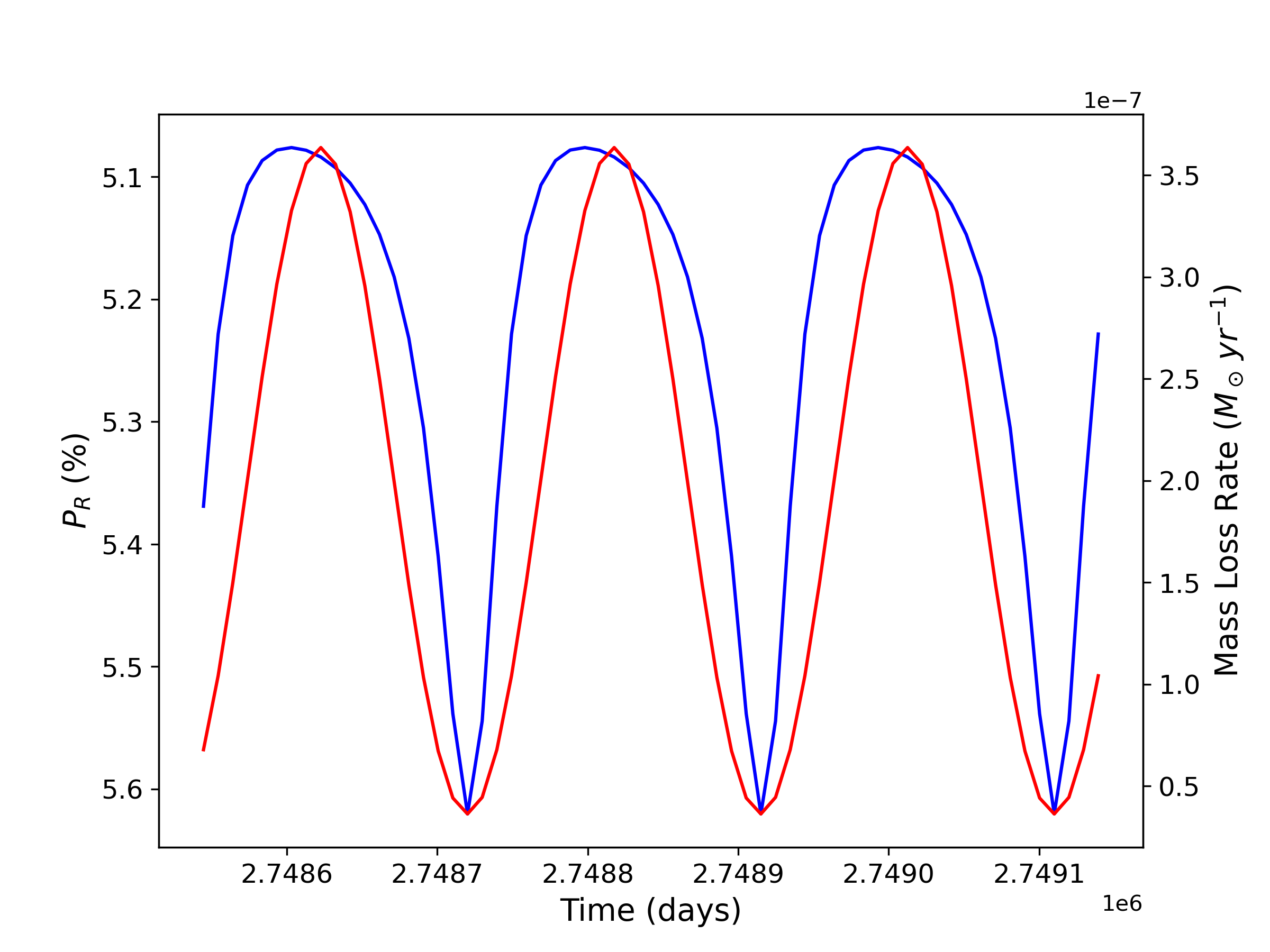}
    \caption{The residual polarization calculated by an approximate Mie scattering theory (blue) and the mass-loss rate (red) as a function of time. The time coordinates for this plot are the nearest 80 points to the time slice $t=4.3750\times 10^{11}\,\text{s}$.}
    \label{fig:Mie_Scattering}
\end{figure}
Below the slice plots is a plot of the polarization as a percentage per optical depth and the star's mass-loss rate, which acts as a proxy for its brightness. The polarization increases downward while the mass-loss rate increases upward. This is to help show the inverse relationship between the two curves. Notably, even as time passes, the star's polarization and mass-loss rate exhibit an inverse relationship. The reason for this inverse relationship comes from the structure of the density data. One may notice a few details by examining the density heat maps and the density from the slice plots. The material in the rarefaction fan nearest the wind bubble has extremely high density in all plots. This is because the star has a large mass-loss rate with a low wind speed. Looking at the heat map, one may notice that this region, especially nearest the wind bubble, is quite spherical. Thus, in calculating the polarization, this region, having a high density and spherical shape, will tend to decrease the polarization signal \citep{Brown1977}, pushing it toward a vanishing spherical value. Moving further from the wind bubble boundary, yet staying in the rarefied region both in the direction of motion and in the tail, the density continues to decrease, and this region has a prolate shape, meaning it is elongated along the z-axis, the direction of motion. This manifests as an asymmetry that contributes to a polarization signal. Next, one encounters the reverse shock, which compresses the stellar wind material and causes the density to be higher than in some parts of the rarefaction fan.
\begin{figure}[t!]
    \centering
    \includegraphics[width=\linewidth]{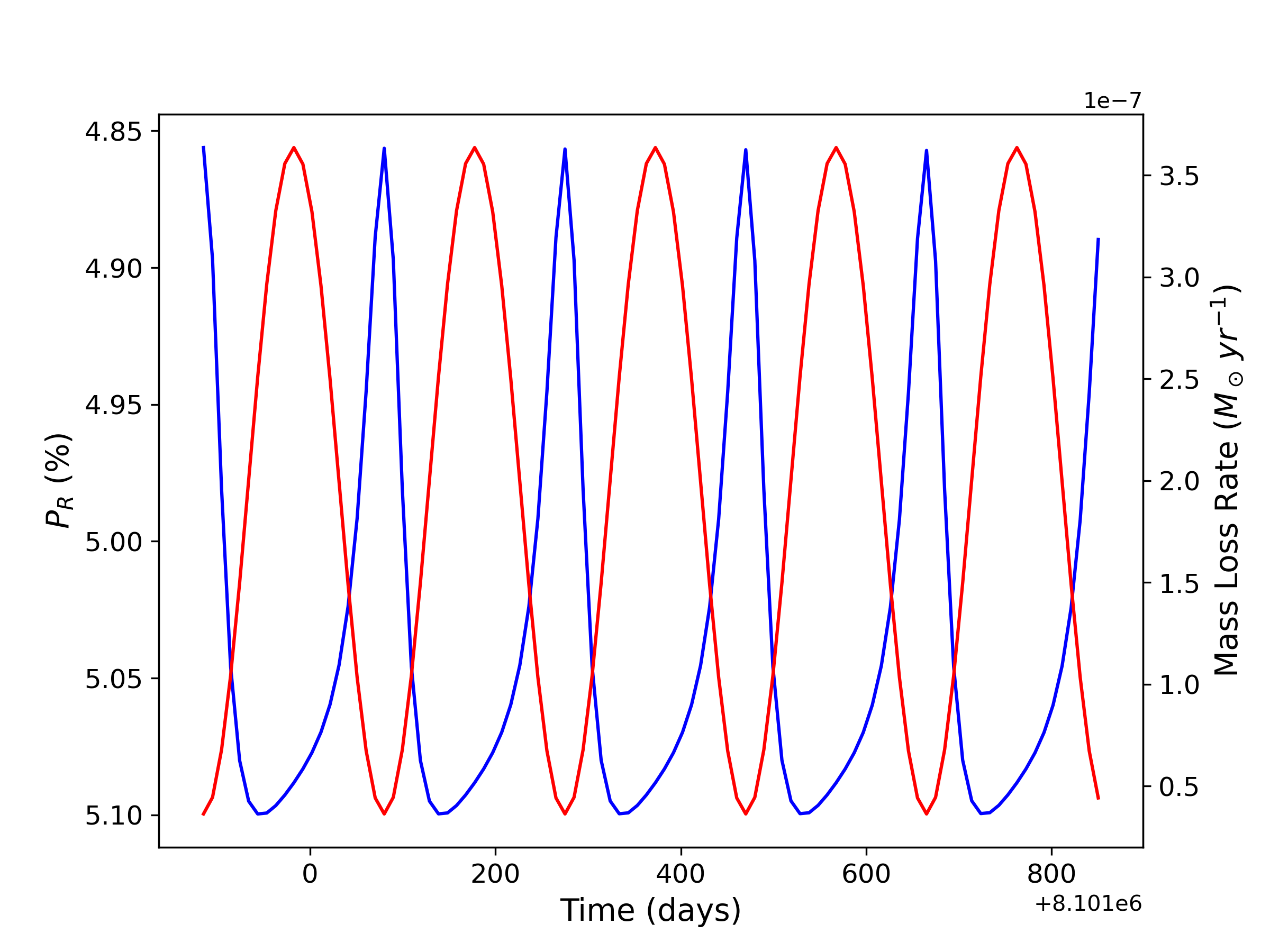}
    \caption{The residual polarization calculated by an approximate Mie scattering theory (blue) and the mass-loss rate (red) as a function of time. The time coordinates for this plot are the nearest 80 points to the time slice $t=7.00\times 10^{11}\,\text{s}$.}
    \label{fig:Mie_Scattering_2}
\end{figure}
Since a bow shock is, by nature, an asymmetric structure, this shocked region contributes the majority of the asymmetry in the density profile and thereby contributes the highest polarization signal, pushing it away from zero. Also of interest is the cometary structure of the tail, which has a shock structure that varies with time in all cases. Due to this, the tail has non-negligible density, contributing to the polarization in a capacity that depends upon the shock structure at the given time. One may picture the manifestation of the inverse relationship in the following way. The material making up the bow shock in the region after the reverse shock sustains its highly compressed value with time. As mentioned, since the bow shock is an asymmetric structure, this contributes the largest polarization signal. While the bow shock is asymmetric, the material in the rarefaction fan near the wind bubble is highly symmetric.

Since the material in the rarefied region has a much higher density, it dominates the integral, pushing the polarization signal toward zero. However, the density coming from the wind bubble varies with time due to the mass loss of the star varying with time. Thus, when the mass-loss rate is at a maximum, so too is the density of the wind. Therefore, the density in the rarefaction fan almost entirely takes over the integral, pushing the polarization toward zero, thus causing it to attain a minimum. However, as the mass-loss rate approaches a minimum, the contribution of the region near the star, while still symmetric, does not dominate the integral as much since the density is lower and the value of the density in the compressed region after the reverse shock stays relatively stable. This causes the asymmetric bow shock to dominate more in the integral, causing a maximum polarization signal at minimum mass loss. Whenever the mass loss is at a maximum, the polarization is roughly at a minimum. Whenever the mass loss is at a minimum, the polarization is roughly at a maximum. Although the physics of the current simulations have been very basic, and the polarization being simple Thomson scattering, or averaged Mie scattering, this is an excellent numerical demonstration of how the variable wind bow shock hypothesis works in practice to give the inverse relationship observed between the signals of V CVn. 

Another aspect of dynamic interest in this interaction is the onset of a Rayleigh--Taylor instability (R-TI), which develops at the interface of the contact discontinuity located at the apex of the bow shock. In 1950, \cite{taylor1950instability} discovered that the instability found by Lord Rayleigh \citep{rayleigh1879dynamical} involving the interface between a fluid of higher density sitting on top of a fluid of lower density in a gravitational field is far more general and happens under any acceleration, not just gravitational \citep{Sharp1984}. According to \cite{drazin2002introduction}, ``there is instability if and only if the net acceleration is directed from the lighter toward the heavier fluid." The R-TI begins with a perturbation in the density profile, likely caused here by numerical viscosity. It then grows as `fingers' in the linear regime of the instability and quickly grows into `mushroom-type clouds', which are susceptible to the Kelvin--Helmholtz instability. In the case of the simulations presented here in this work, a net acceleration is caused by the material passing through the forward shock and decelerating the contact discontinuity at the apex of the bow shock. Due to the high density of the stellar wind being swept up by the reverse shock, the left-hand side of the contact discontinuity (closer to the star) is far denser than the right-hand side of the contact discontinuity (the ISM material swept up by the forward shock). Since the shocked ISM material is causing a deceleration to the contact discontinuity, this is a net acceleration pointing from the direction of the much less dense shocked ISM to the much more dense shocked stellar wind across the contact. The numerical viscosity then initiates the R-TI in the contact discontinuity, which first acts as fingers. When the mushroom-type structures begin to form, which are Kelvin-Helmholtz unstable in the non-linear regime of the R-TI, they don't have enough time to grow too much because they are swept into the shear flow of the shocked ISM material making its way into the tail. However, this behaviour, along with the mixing of the `fingers', can cause large perturbations to grow at the interface of the contact discontinuity, leading to the tail destabilizing slightly and changing the morphology. This can be seen toward the end of the constant stellar wind velocity simulation, Figure \ref{12}. This result should be taken cautiously, as these simulations are only two-dimensional. Full three-dimensional simulations should be examined if one wishes to probe if this instability is truly real. There is some precedent for the onset of this R-TI in the literature, in the work of \cite{Villaver2012}, who used the other version of Zeus3D from \cite{Stone1992ApJS...80..753S}, as well as \cite{Meyer2014}. Lastly, another interesting aspect of the results is the temperature structure in the tail. Due to the shock heating, the tail attains an extremely high temperature. This indicates numerical evidence of a ``Mira-like tail" \citep{Neilson2014}.

While the polarization and mass-loss rate plots at various short time slices allow one to gain an understanding of the relationship between the variables, the entire dataset of over $20000$ years cannot be shown in this way on a singular plot due to how quickly the data oscillates compared to the total period. Therefore, another method of displaying the entire dataset coupled with the plots is required to understand the relationship fully. The most useful and intuitive process for showing the relationship between the variables in the whole dataset is a normalized cross-correlation function. The definition of a cross-correlation function \citep{press2003numerical} is
\begin{equation}
    C(\tau_\text{l}) = \int_{-\infty}^{\infty} P_\text{R}(t)\dot{M}(t+\tau_\text{l})\, dt.
\end{equation}
Here, $\tau_\text{l}$ represents the lag. For this work, the normalization of the cross-correlation function has a mean that vanishes and a standard deviation of unity. This makes the interpretation of the function easier to understand. The dataset has a strong linear relationship when the normalized cross-correlation function has a strong positive value. Conversely, the dataset has a strong inverse relationship when the normalized cross-correlation function has a strong negative value. Weak or vanishing values imply little or no correlation in the dataset. Mathematically, 
\begin{equation}
    \hat{C}(\tau_\text{l}) = \frac{C(\tau_\text{l}) - \mu(C(\tau_\text{l}))}{\sigma(C(\tau_\text{l}))}.
\end{equation}
Here, $\hat{C}(\tau_\text{l})$ is the normalized cross-correlation function, $\mu(C(\tau_\text{l}))$ represents the mean of the cross-correlation function, and $\sigma(C(\tau_\text{l}))$ represents the standard deviation of the cross-correlation function. One may show, using Figure \ref{Cross_Corr_Static}, that the normalized cross-correlation function of V CVn has a strong negative peak near zero, and at zero lag, they also have a strong negative value. This implies that the function has a strong inverse relationship across the entire dataset due to the mean and standard deviation of $\hat{C}(\tau_\text{l})$. However, one may notice that the cross-correlation function does not peak at the origin; rather, it peaks in the negative direction near the origin. Thus, while the actual data is strongly inversely related as is, it is not naturally maximally inversely related. Instead, the polarization leads the mass-loss rate by a small number of days for each model. 
%%%%% Figures %%%%%
%%%%% End Static %%%%%
%%%%% End Variable %%%%%

\subsection{Approximate Mie Scattering Results}\label{sec:Approximate_Mie}
While the main focus of this research has been a simplistic treatment using Thomson scattering to develop an understanding of underlying physics and mechanics, a more realistic treatment of scattered light from V CVn involves dust grains. In Section \ref{sec:Approximate_Mie}, we developed an effective Mie scattering theory by assuming averaged quantities. Figure \ref{fig:Mie_Scattering} is the residual polarization $P_\text{R}$ and the mass-loss rate plotted against time. Even with a very approximate theory of Mie scattering, we find that the roughly inverse relationship between brightness and polarization is maintained, and the polarization signal attains levels comparable to the photometric observations presented in \cite{Neilson2014} and \cite{Neilson2023}. The Mie scattering results of Figure \ref{fig:Mie_Scattering_2} indicate that the inverse relationship of polarization and mass-loss rate is destroyed at certain times. In fact, the relationship goes back to what one would expect, which is highly linearly correlated. It seems that this change from inverse correlation to linear correlation occurs around times when instabilities tear apart the bow shock. We leave further investigation of this phenomenon for future three-dimensional simulations with the ability to have the polarization position angle as a diagnostic tool. 

\section{Conclusions}\label{sec:Conclusions}
The results of both numerical models for the wind of the star V CVn show that the polarization signal produced varies inversely with its mass-loss rate (a proxy for the star's brightness). Observations of V CVn show that polarization and brightness are roughly inversely correlated. Therefore, this work provides strong numerical evidence that when the mass loss of the star is at its maximum, the envelope around the star is dense, meaning it dominates the polarization integrals, and quite symmetric, meaning it tends to push the polarization signal to zero causing it to attain a minimum. Conversely, when the mass-loss rate is at a minimum, the stellar wind material and the compressed ISM material in the post-shock regions of the asymmetric bow shock begin to contribute much more to the polarization integral in comparison to the now less-dense symmetric shell around the star at a minimum mass-loss rate. Therefore, the asymmetric structure of the bow shock pushes the polarization signal to a maximum. Thus, it is plausible that the curious case of V CVn is solved by the existence of an asymmetric stellar wind bow shock driven by a dusty, pulsating wind.

Moving forward, there are many possibilities for continuing this work. The first and most obvious is full three-dimensional simulations. The problem with needing to simulate a bow shock for tens of thousands of years, driven by a wind which varies on the order of two hundred days, is that the wind evolution requires a time step small enough to resolve it adequately. In three dimensions, this constraint severely increases the computational cost. However, three-dimensional simulations immediately give access to several pieces of information. Most important is the calculation of the full suite of Stokes parameters. Having all Stokes parameters changes the polarization to $P_\text{R}=\sqrt{Q^2+U^2}/I_\text{tot}$, but now gives access to information about the polarization position angle defined by $\tan(2\psi)=U/Q$. The polarization is not the only strange behaviour of V CVn; it is also the near-constant position angle. In three dimensions, asymmetries in the envelope caused by instabilities and turbulence will manifest as a variation in the polarization position angle. However, these variations will likely be quite small, and therefore, one may find a nearly constant position angle and contribute more evidence to the variable wind bow shock hypothesis. 
As demonstrated in Section \ref{sec:Approximate_Mie}, more realistic physics should be included, such as a better accounting of Mie scattering, by either adding a routine to calculate it in \textsc{ZEUS3D} on-the-fly or using the established framework to export the time variation of the density field for use by the code \textsc{SLIP} \citep{hoffman2007, Shrestha2018}. Mie scattering is severely more realistic, and as we've shown, even in an approximate case, it clearly helps amplify the polarization signal coming from the bow shock, which causes the signal to become closer to the observed strength. Moreover, one could also include the effects of multiple scattering instead of single scattering, and finally, a proper stellar wind model could be added to \textsc{ZEUS3D} to govern the time variation of the wind bubble boundary conditions. 

\section*{Acknowledgements}

The authors thank the people and entities who made this research possible. We thank Dr David A. Clarke for his illuminating discussions on computational magnetohydrodynamics and Dr Christopher D. Matzner for his incredibly useful comments on the manuscript. Also, the authors would like to thank the National Sciences and Engineering Research Council of Canada for funding provided by a Canada Graduate Scholarship-Master’s (CGS M) and Memorial University of Newfoundland for providing funding via a Dean's Scholarship. Finally, the authors would like to thank Compute Ontario and the Digital Research Alliance of Canada for the use of the Niagara supercomputer.

This work has made use of data from the European Space Agency (ESA) mission
{\it Gaia} (\url{https://www.cosmos.esa.int/gaia}), processed by the {\it Gaia}
Data Processing and Analysis Consortium (DPAC,
\url{https://www.cosmos.esa.int/web/gaia/dpac/consortium}). Funding for the DPAC
has been provided by national institutions, in particular, the institutions
participating in the {\it Gaia} Multilateral Agreement.
%%%%%%%%%%%%%%%%%%%%%%%%%%%%%%%%%%%%%%%%%%%%%%%%%%
\section*{Data Availability}
All data are available upon reasonable request.
%%%%%%%%%%%%%%%%% APPENDICES %%%%%%%%%%%%%%%%%%%%%

%\bibliography{VCVN}{}

\begin{thebibliography}{}
\expandafter\ifx\csname natexlab\endcsname\relax\def\natexlab#1{#1}\fi
\providecommand{\url}[1]{\href{#1}{#1}}
\providecommand{\dodoi}[1]{doi:~\href{http://doi.org/#1}{\nolinkurl{#1}}}
\providecommand{\doeprint}[1]{\href{http://ascl.net/#1}{\nolinkurl{http://ascl.net/#1}}}
\providecommand{\doarXiv}[1]{\href{https://arxiv.org/abs/#1}{\nolinkurl{https://arxiv.org/abs/#1}}}

% type= book
\bibitem[{G.~B. Arfken {et~al.}(2012)Arfken, Weber, \& Harris}]{arfken2012mathematical}
Arfken, G.~B., Weber, H.~J., \& Harris, F.~E. 2012, Mathematical Methods for Physicists: A Comprehensive Guide, 7th edn. (London: Academic Press)

% type= article
\bibitem[{C. {Babusiaux} {et~al.}(2023){Babusiaux}, {Fabricius}, {Khanna}, {et~al.}}]{2023A&A...674A..32B}
{Babusiaux}, C., {Fabricius}, C., {Khanna}, S., {et~al.} 2023, \bibinfo{title}{{Gaia Data Release 3. Catalogue validation},} Astronomy \& Astrophysics, 674, A32, \dodoi{10.1051/0004-6361/202243790}

% type= article
\bibitem[{V.~B. Baranov {et~al.}(1971)Baranov, Krasnobaev, \& Kulikovskii}]{baranov1971}
Baranov, V.~B., Krasnobaev, K.~V., \& Kulikovskii, A.~G. 1971, \bibinfo{title}{Soviet Phys.–Dokl.,} Soviet Physics--Doklady, 15, 791

% type= book
\bibitem[{W.~H. Beyer(1987)Beyer}]{Beyer1987}
Beyer, W.~H. 1987, CRC Standard Mathematical Tables, 28th edn. (Boca Raton, FL: CRC Press)

% type= article
\bibitem[{J.~C. Brown \& I.~S. McLean(1977)Brown \& McLean}]{Brown1977}
Brown, J.~C., \& McLean, I.~S. 1977, \bibinfo{title}{Polarisation by Thomson Scattering in Optically Thin Stellar Envelopes. I. Source Star at Centre of Axisymmetric Envelope,} Astronomy and Astrophysics, 57, 141

% type= article
\bibitem[{J.~C. Brown {et~al.}(1978)Brown, McLean, \& Emslie}]{Brown1978}
Brown, J.~C., McLean, I.~S., \& Emslie, A.~G. 1978, \bibinfo{title}{Polarisation by Thomson scattering in optically thin stellar envelopes. II. Binary and multiple star envelopes and the determination of binary inclinations.,} Astronomy and Astrophysics, 68, 415

% type= book
\bibitem[{S. Chandrasekhar(1960)Chandrasekhar}]{Chandrasekhar1960}
Chandrasekhar, S. 1960, Radiative Transfer (New York: Dover Publications)

% type= article
\bibitem[{D.~A. {Clarke}(1996){Clarke}}]{Clarke1996ApJ...457..291C}
{Clarke}, D.~A. 1996, \bibinfo{title}{{A Consistent Method of Characteristics for Multidimensional Magnetohydrodynamics},} \apj, 457, 291, \dodoi{10.1086/176730}

% type= article
\bibitem[{D.~A. Clarke(2010)Clarke}]{Clarke_2010}
Clarke, D.~A. 2010, \bibinfo{title}{ON THE RELIABILITY OF ZEUS-3D,} The Astrophysical Journal Supplement Series, 187, 119, \dodoi{10.1088/0067-0049/187/1/119}

% type= manual
\bibitem[{D.~A. Clarke(2016)Clarke}]{clarke2016zeus3d}
Clarke, D.~A. 2016, What is ZEUS-3D?, Institute for Computational Astrophysics, Saint Mary's University, Halifax, NS, Canada.
\newblock \url{http://www.ica.smu.ca/zeus3d}

% type= article
\bibitem[{ {De Beck, E.} {et~al.}(2010){De Beck, E.}, {Decin, L.}, {de Koter, A.}, {Justtanont, K.}, {Verhoelst, T.}, {Kemper, F.}, \& {Menten, K. M.}}]{DeBeck}
{De Beck, E.}, {Decin, L.}, {de Koter, A.}, {et~al.} 2010, \bibinfo{title}{Probing the mass-loss history of AGB and red supergiant stars from CO rotational line profiles* - II. CO line survey of evolved stars: derivation of mass-loss rate formulae,} A\&A, 523, A18, \dodoi{10.1051/0004-6361/200913771}

% type= book
\bibitem[{B.~T. {Draine}(2011){Draine}}]{Draine_Book}
{Draine}, B.~T. 2011, {Physics of the Interstellar and Intergalactic Medium} (Princeton University Press)

% type= book
\bibitem[{P.~G. Drazin(2002)Drazin}]{drazin2002introduction}
Drazin, P.~G. 2002, Introduction to Hydrodynamic Stability (Cambridge: Cambridge University Press)

% type= article
\bibitem[{B. Famaey {et~al.}(2009)Famaey {et~al.}}]{2009A&A...498..627F}
Famaey, B., {et~al.} 2009, \bibinfo{title}{Kinematics of local AGB stars,} Astronomy \& Astrophysics, 498, 627

% type= misc
\bibitem[{ {Gaia Collaboration}(2020){Gaia Collaboration}}]{2020yCat.1350....0G}
{Gaia Collaboration}. 2020, Gaia Data Release 2 Catalogue,, \url{https://vizier.u-strasbg.fr/viz-bin/VizieR?-source=I/345}

% type= article
\bibitem[{ {Gaia Collaboration} {et~al.}(2016){Gaia Collaboration}, {Prusti}, {de Bruijne}, {Brown}, {Vallenari}, {Babusiaux}, {Bailer-Jones}, {et~al.}}]{2016A&A...595A...1G}
{Gaia Collaboration}, {Prusti}, T., {de Bruijne}, J.~H.~J., {et~al.} 2016, \bibinfo{title}{{The Gaia mission},} Astronomy \& Astrophysics, 595, A1, \dodoi{10.1051/0004-6361/201629272}

% type= article
\bibitem[{ {Gaia Collaboration} {et~al.}(2023){Gaia Collaboration}, {Vallenari}, {Brown}, {Prusti}, {de Bruijne}, {et~al.}}]{2023A&A...674A...1G}
{Gaia Collaboration}, {Vallenari}, A., {Brown}, A.~G.~A., {et~al.} 2023, \bibinfo{title}{{Gaia Data Release 3. Summary of the content and survey properties},} Astronomy \& Astrophysics, 674, A1, \dodoi{10.1051/0004-6361/202243940}

% type= book
\bibitem[{I.~S. Gradshteyn \& I.~M. Ryzhik(2014)Gradshteyn \& Ryzhik}]{GradshteynRyzhik2014}
Gradshteyn, I.~S., \& Ryzhik, I.~M. 2014, Table of Integrals, Series, and Products, 8th edn. (Amsterdam: Academic Press), \dodoi{10.1016/C2010-0-64839-5}

% type= book
\bibitem[{J.~G. L. M.~L. Henny \& J.~P. Cassinelli(1999)Henny \& Cassinelli}]{henny_stellar_winds}
Henny, J. G. L. M.~L., \& Cassinelli, J.~P. 1999, Introduction to Stellar Winds (Cambridge: Cambridge University Press), \dodoi{10.1017/CBO9781139175012}

% type= inproceedings
\bibitem[{J.~L. Hoffman(2007)Hoffman}]{hoffman2007}
Hoffman, J.~L. 2007, \bibinfo{title}{SLIP: A Monte Carlo Radiation Transfer Code,} in Revista Mexicana de Astronom{\'\i}a y Astrof{\'\i}sica Conference Series, Vol.~30, Circum-stellar Media and Late Stages of Massive Stellar Evolution, ed. G.~Garc{\'\i}a-Segura \& E.~Ram{\'\i}rez-Ruiz, Mexico, 57

% type= article
\bibitem[{S. H{\"o}fner \& H. Olofsson(2018)H{\"o}fner \& Olofsson}]{Hofner2018}
H{\"o}fner, S., \& Olofsson, H. 2018, \bibinfo{title}{Mass loss of stars on the asymptotic giant branch,} Astronomy and Astrophysics Review, 26, \dodoi{10.1007/s00159-017-0106-5}

% type= article
\bibitem[{H.~L. Johnson \& W.~W. Morgan(1937)Johnson \& Morgan}]{1937ApJ....85....9J}
Johnson, H.~L., \& Morgan, W.~W. 1937, \bibinfo{title}{Spectral Classification of M-type Stars,} The Astrophysical Journal, 85, 9

% type= article
\bibitem[{J. Mackey {et~al.}(2021)Mackey, Green, Moutzouri, Haworth, Kavanagh, Zargaryan, \& Celeste}]{MacKey2021}
Mackey, J., Green, S., Moutzouri, M., {et~al.} 2021, \bibinfo{title}{Pion: Simulating bow shocks and circumstellar nebulae,} Monthly Notices of the Royal Astronomical Society, 504, 983, \dodoi{10.1093/mnras/stab781}

% type= article
\bibitem[{A.~M. Magalhães {et~al.}(1986)Magalhães, Coyne, \& Benedetti}]{Magalhaes1986}
Magalhães, A.~M., Coyne, G.~V., \& Benedetti, E.~K. 1986, \bibinfo{title}{Polarimetry of Stars with Infrared Excesses. II,} The Astronomical Journal, 91, 919, \dodoi{10.1086/114064}

% type= article
\bibitem[{D.~M. Meyer {et~al.}(2014)Meyer, Gvaramadze, Langer, Mackey, Boumis, \& Mohamed}]{Meyer2014}
Meyer, D.~M., Gvaramadze, V.~V., Langer, N., {et~al.} 2014, \bibinfo{title}{On the stability of bow shocks generated by red supergiants: The case of IRC -10414,} Monthly Notices of the Royal Astronomical Society: Letters, 439, \dodoi{10.1093/mnrasl/slt176}

% type= article
\bibitem[{H. Neilson {et~al.}(2023)Neilson, Steenken, Simpson, Ignace, Shrestha, Erba, \& Henson}]{Neilson2023}
Neilson, H., Steenken, N., Simpson, J., {et~al.} 2023, \bibinfo{title}{A multiyear photopolarimetric study of the semi-regular variable V CVn and identification of analog sources,} Astronomy and Astrophysics, 677, \dodoi{10.1051/0004-6361/202245154}

% type= article
\bibitem[{H.~R. Neilson {et~al.}(2014)Neilson, Ignace, Smith, Henson, \& Adams}]{Neilson2014}
Neilson, H.~R., Ignace, R., Smith, B.~J., Henson, G., \& Adams, A.~M. 2014, \bibinfo{title}{Evidence of a Mira-like tail and bow shock about the semi-regular variable v CVn from four decades of polarization measurements,} Astronomy and Astrophysics, 568, \dodoi{10.1051/0004-6361/201424037}

% type= article
\bibitem[{T.~A. Poliakova(1981)Poliakova}]{Poliakova1981}
Poliakova, T.~A. 1981, \bibinfo{title}{Study on the Structure of Stellar Atmospheres,} Leningradskii Universitet Vestnik Matematika Mekhanika Astronomiia, 2, 105

% type= book
\bibitem[{W.~H. Press {et~al.}(2003)Press, Teukolsky, Vetterling, \& Flannery}]{press2003numerical}
Press, W.~H., Teukolsky, S.~A., Vetterling, W.~T., \& Flannery, B.~P. 2003, Numerical Recipes in Fortran 77: The Art of Scientific Computing, 2nd edn. (Cambridge: Cambridge University Press)

% type= article
\bibitem[{L. Rayleigh(1879)Rayleigh}]{rayleigh1879dynamical}
Rayleigh, L. 1879, \bibinfo{title}{On the Dynamical Theory of Gravitation,} Proceedings of the Royal Society of London, 10, 170

% type= article
\bibitem[{L.~F. Richardson(1911)Richardson}]{richardson1911approximate}
Richardson, L.~F. 1911, \bibinfo{title}{The approximate arithmetical solution by finite differences of physical problems involving differential equations, with an application to the stresses in a masonry dam,} Philosophical Transactions of the Royal Society of London. Series A, Containing Papers of a Mathematical or Physical Character, 210, 307, \dodoi{10.1098/rsta.1911.0009}

% type= article
\bibitem[{L.~F. Richardson \& J.~A. Gaunt(1927)Richardson \& Gaunt}]{richardson1927deferred}
Richardson, L.~F., \& Gaunt, J.~A. 1927, \bibinfo{title}{The deferred approach to the limit,} Philosophical Transactions of the Royal Society of London. Series A, Containing Papers of a Mathematical or Physical Character, 226, 299, \dodoi{10.1098/rsta.1927.0008}

% type= article
\bibitem[{F. {Rouleau} \& P.~G. {Martin}(1991){Rouleau} \& {Martin}}]{1991ApJ...377..526R}
{Rouleau}, F., \& {Martin}, P.~G. 1991, \bibinfo{title}{{Shape and Clustering Effects on the Optical Properties of Amorphous Carbon},} \apj, 377, 526, \dodoi{10.1086/170382}

% type= article
\bibitem[{B.~S. Safonov {et~al.}(2019)Safonov, Dodin, Lamzin, \& Rastorguev}]{Safonov2019}
Safonov, B.~S., Dodin, A.~V., Lamzin, S.~A., \& Rastorguev, A.~S. 2019, \bibinfo{title}{The Circumstellar Envelope of the Semiregular Variable Star V CVn,} Astronomy Letters, 45, 453, \dodoi{10.1134/S1063773719070065}

% type= article
\bibitem[{D.~H. Sharp(1984)Sharp}]{Sharp1984}
Sharp, D.~H. 1984, \bibinfo{title}{An overview of Rayleigh-Taylor instability,} Physica D: Nonlinear Phenomena, 12, 3, \dodoi{10.1016/0167-2789(84)90510-4}

% type= article
\bibitem[{M. Shrestha {et~al.}(2018)Shrestha, Neilson, Hoffman, \& Ignace}]{Shrestha2018}
Shrestha, M., Neilson, H.~R., Hoffman, J.~L., \& Ignace, R. 2018, \bibinfo{title}{Polarization simulations of stellar wind bow-shock nebulae - I. The case of electron scattering,} Monthly Notices of the Royal Astronomical Society, 477, 1365, \dodoi{10.1093/mnras/sty724}

% type= article
\bibitem[{M. Shrestha {et~al.}(2021)Shrestha, Neilson, Hoffman, Ignace, \& Fullard}]{Shrestha2021}
Shrestha, M., Neilson, H.~R., Hoffman, J.~L., Ignace, R., \& Fullard, A.~G. 2021, \bibinfo{title}{Polarization simulations of stellar wind bow shock nebulae - II. The case of dust scattering,} Monthly Notices of the Royal Astronomical Society, 500, 4319, \dodoi{10.1093/mnras/staa3508}

% type= article
\bibitem[{J.~M. {Stone} \& M.~L. {Norman}(1992){Stone} \& {Norman}}]{Stone1992ApJS...80..753S}
{Stone}, J.~M., \& {Norman}, M.~L. 1992, \bibinfo{title}{{ZEUS-2D: A Radiation Magnetohydrodynamics Code for Astrophysical Flows in Two Space Dimensions. I. The Hydrodynamic Algorithms and Tests},} \apjs, 80, 753, \dodoi{10.1086/191680}

% type= article
\bibitem[{B.~J. Sumlin {et~al.}(2018)Sumlin, Heinson, \& Chakrabarty}]{SUMLIN2018127}
Sumlin, B.~J., Heinson, W.~R., \& Chakrabarty, R.~K. 2018, \bibinfo{title}{Retrieving the aerosol complex refractive index using PyMieScatt: A Mie computational package with visualization capabilities,} Journal of Quantitative Spectroscopy and Radiative Transfer, 205, 127, \dodoi{https://doi.org/10.1016/j.jqsrt.2017.10.012}

% type= article
\bibitem[{G.~I. Taylor(1950)Taylor}]{taylor1950instability}
Taylor, G.~I. 1950, \bibinfo{title}{The Instability of Liquid Surfaces When Accelerated in a Direction Perpendicular to Their Planes,} Proceedings of the Royal Society of London. Series A, Mathematical and Physical Sciences, 201, 192

% type= article
\bibitem[{E. Villaver {et~al.}(2012)Villaver, Manchado, \& García-Segura}]{Villaver2012}
Villaver, E., Manchado, A., \& García-Segura, G. 2012, \bibinfo{title}{The interaction of asymptotic giant branch stars with the interstellar medium,} Astrophysical Journal, 748, \dodoi{10.1088/0004-637X/748/2/94}

% type= article
\bibitem[{J. {Von Neumann} \& R.~D. {Richtmyer}(1950){Von Neumann} \& {Richtmyer}}]{VonNeumann1950}
{Von Neumann}, J., \& {Richtmyer}, R.~D. 1950, \bibinfo{title}{{A Method for the Numerical Calculation of Hydrodynamic Shocks},} Journal of Applied Physics, 21, 232, \dodoi{10.1063/1.1699639}

% type= article
\bibitem[{M. Wenger {et~al.}(2000)Wenger, Ochsenbein, Egret, Dubois, Bonnarel, Borde, Genova, Jasniewicz, Laloë, Lesteven, \& Monier}]{Wenger_2000}
Wenger, M., Ochsenbein, F., Egret, D., {et~al.} 2000, \bibinfo{title}{The SIMBAD astronomical database: The CDS reference database for astronomical objects,} Astronomy and Astrophysics Supplement Series, 143, 9–22, \dodoi{10.1051/aas:2000332}

% type= article
\bibitem[{F.~P. Wilkin(1996)Wilkin}]{Wilkin1996}
Wilkin, F.~P. 1996, \bibinfo{title}{EXACT ANALYTIC SOLUTIONS FOR STELLAR WIND BOW SHOCKS,} The Astrophysical Journal, 459, L31

% type= article
\bibitem[{J.~M. Winters {et~al.}(2000)Winters, Le~Bertre, Jeong, Helling, \& Sedlmayr}]{Winters2000}
Winters, J.~M., Le~Bertre, T., Jeong, K.~S., Helling, C., \& Sedlmayr, E. 2000, \bibinfo{title}{A systematic investigation of the mass loss mechanism in dust forming long-period variable stars,} Astronomy and Astrophysics, 361, 641

% type= article
\bibitem[{M.~J. Wolff {et~al.}(1996)Wolff, Nordsieck, \& Nook}]{Wolff1996}
Wolff, M.~J., Nordsieck, K.~H., \& Nook, M.~A. 1996, \bibinfo{title}{An Ultraviolet Interstellar Polarization Survey: Stars with High Color Excess,} The Astronomical Journal, 111, 856, \dodoi{10.1086/117822}

\end{thebibliography}
%\bibliographystyle{aasjournalv7}

\end{document}